\documentclass[trackchanges, twocolumn]{aastex701}
\usepackage{amsmath}
\usepackage{amssymb}
\usepackage{amsfonts}
\usepackage{times}
\usepackage{color}
\usepackage{subcaption}
\usepackage{soul}
\usepackage{enumerate}
\usepackage{enumitem}
\usepackage{xcolor}

\renewcommand{\d}{\mathrm{d}}

\begin{document}

\title{Model-Independent Indication for a Localized Anomaly in the Late-Time Expansion History}

\author[orcid=0000-0002-3840-9456,sname='Gupta Choudhury']{Shibendu Gupta Choudhury}
\email[show]{pdf.schoudhury@jmi.ac.in}
\affiliation{Centre for Theoretical Physics, Jamia Millia Islamia, New Delhi 110025, India}
\affiliation{Indian Institute of Astrophysics, Bengaluru, Karnataka 560034, India}

\author[orcid=0000-0002-2701-5654,sname='Mukherjee']{Purba Mukherjee}
\affiliation{Centre for Theoretical Physics, Jamia Millia Islamia, New Delhi 110025, India}
\affiliation{Korea Astronomy and Space Science Institute (KASI), Daejeon 34055, Republic of Korea}
\email[show]{pdf.pmukherjee@jmi.ac.in}

\author[orcid=0000-0001-8408-6961,sname='Di Valentino']{Eleonora Di Valentino}
\email[show]{e.divalentino@sheffield.ac.uk}
\affiliation{School of Mathematical and Physical Sciences, University of Sheffield, Hounsfield Road, Sheffield S3 7RH, United Kingdom}

\author[orcid=0000-0001-9615-4909,sname='Sen']{Anjan A Sen}
\affiliation{Centre for Theoretical Physics, Jamia Millia Islamia, New Delhi 110025, India}
\email[show]{aasen@jmi.ac.in}

\begin{abstract}
We investigate the late-time expansion history of the Universe using a model-independent spline reconstruction of cosmological distances based on the latest DESI DR2 baryon acoustic oscillation (BAO) measurements and the DES Dovekie Type Ia supernova compilation. Comparing the reconstructed expansion history with the prediction of the Planck 2018 $\Lambda$CDM model, we identify a localized deviation over the redshift interval $0.3\lesssim z\lesssim0.6$, reaching a maximum significance of approximately $3.5\sigma$ at $z\simeq0.47$. We demonstrate that this feature persists under substantial variations of the reconstruction methodology, dataset composition and sound-horizon calibration.
Mock analyses further show that the reconstruction is unbiased and that the observed anomaly is unlikely to arise from reconstruction bias or miscalibrated uncertainties. If confirmed by future observations, this localized feature could point to previously unrecognized late-time physics or reveal subtle inconsistencies between early and late Universe cosmological probes.
\end{abstract}


\section{Introduction}

Understanding the origin of the accelerated expansion of the Universe remains one of the central challenges of modern cosmology. Although the concordance $\Lambda$CDM model successfully describes a broad range of observations, several recent results have motivated continued investigations into possible departures from the standard cosmological framework. These include the persistent Hubble tension and $S_8/\sigma_8$ discrepancy~\citep{Riess:2021jrx,DiValentino:2021izs,Abdalla:2022yfr,H0DN:2025lyy,CosmoVerseNetwork:2025alb,DiValentino:2020vvd,DES:2026fyc,Wright:2025xka}, the unexpectedly large abundance of massive galaxies at high redshift reported by the James Webb Space Telescope (JWST)~\citep{Haslbauer:2022vnq,Lovell:2022bhx,Boylan-Kolchin:2022kae}, and recent analyses combining Dark Energy Spectroscopic Instrument (DESI) BAO with Type Ia supernova (SNIa) and cosmic microwave background (CMB) observations, which have provided evidence for departures from a cosmological constant at the $\sim3.2$--$3.4\sigma$ level, depending on the dataset combination~\citep{DESI:2024mwx,DESI:2025zgx,Scolnic:2021amr,Rubin:2023jdq,DES:2025sig,Planck:2018vyg,ACT:2025fju,SPT-3G:2025bzu,Hoyt:2026fve}. With the advent of ongoing and forthcoming high-precision surveys such as LSST~\citep{LSST:2008ijt,LSSTScience:2009jmu,breivik2022datasoftwaresciencerubin} and Euclid~\citep{Euclid:2019clj,Euclid:2021xmh,Euclid:2024yrr,EuclidTheoryWorkingGroup:2012gxx}, it is becoming increasingly important to develop model-independent approaches capable of probing the expansion history of the Universe with minimal theoretical assumptions.

Most investigations of dark energy adopt either physically motivated models, such as scalar-field scenarios or interacting dark-sector models~\citep{Giare:2024ytc,Giare:2024smz,Li:2024qso,Pooya:2024wsq,Halder:2024uao,Silva:2025hxw,Yang:2025uyv,Zhang:2025dwu,Li:2025muv,Li:2025owk,Park:2025fbl,Wolf:2025acj,Toomey:2025xyo,Hossain:2025grx,Shah:2025ayl,Li:2026xaz,Zhai:2026uwr}, or phenomenological parameterizations of quantities including the dark energy equation of state $w_{\rm DE}$~\citep{Najafi:2024qzm,Giare:2024gpk,Giare:2024ocw,Jiang:2024xnu,RoyChoudhury:2024wri,Giare:2024oil,Giare:2025pzu,Kessler:2025kju,RoyChoudhury:2025dhe,Scherer:2025esj,Wolf:2025jlc,Santos:2025wiv,Cheng:2025hug,Ozulker:2025ehg,Li:2025vuh,Lee:2025pzo,Fazzari:2025lzd,Smith:2025icl,Herold:2025hkb,Cheng:2025yue,Gokcen:2026pkq,Ishak:2025cay,Najafi:2026kxs,Yang:2026yaq,Kessler:2026dbi,Lee:2026yzs,Li:2026asg,Giare:2026oti}, energy density $\rho_{\rm DE}$~\citep{DiValentino:2020naf,Dinda:2025iaq,Adil:2023exv,Mukherjee:2025myk,Lee:2026yzs,Specogna:2025guo}, or pressure $p_{\rm DE}$~\citep{Sen:2007gk,Cheng:2025lod}. Complementary to these approaches, numerous parametric and non-parametric reconstruction techniques have been developed to infer the expansion history directly from cosmological observations, reconstructing quantities such as $H(z)$, the scale factor $a(t)$, $w_{\rm DE}$, and $\rho_{\rm DE}$~\citep{Capozziello:2018jya,Dutta:2018vmq,Roy:2022fif,Mukhopadhyay:2024fch,Choudhury:2025bnx,Dutta:2019pio,Berti:2025phi,Dinda:2024ktd,Jiang:2024xnu,Mukherjee:2024ryz,DESI:2025fii,DESI:2025wyn,Fazzari:2025lzd,Ormondroyd:2025exu,Kessler:2026dbi}.

Among these, distance-based reconstruction provides one of the most direct and least assumption-dependent probes of the expansion history. Rather than postulating a specific dark energy or modified gravity model, it reconstructs cosmological distances from observations and infers the Hubble parameter $H(z)$ through their derivatives. Since cosmological distances depend only on the geometry of the background Universe, this approach is sensitive to the total cosmic energy density and pressure without requiring assumptions about the nature of dark energy, the evolution of individual cosmic components, or their conservation equations. Distance-based reconstructions therefore offer a powerful and robust framework for testing the consistency of the standard cosmological model and searching for localized deviations in the expansion history.

Recently,~\citet{Mukherjee:2025ytj} applied a model-independent distance reconstruction to BAO and SNIa observations calibrated using the Planck 2018 sound horizon. They reported discrepancies with the Planck-$\Lambda$CDM prediction at several discrete redshifts below $z\sim1$, suggesting that any departure from the standard cosmological model inferred from late-time distance measurements may be localized rather than global.

In this work, we investigate this question using the latest DESI-DR2 BAO measurements~\citep{DESI:2025zgx} and the DES Dovekie SNIa compilation~\citep{DES:2025sig}. We perform a free-form, knot-based spline reconstruction of cosmological distances and infer the corresponding expansion history across the full redshift range covered by the observations. Rather than evaluating the consistency with Planck-$\Lambda$CDM only at a small number of characteristic redshifts, we reconstruct the complete tension profile as a continuous function of redshift, allowing us to identify both localized anomalies and broader trends in the expansion history. In addition, we introduce a novel physics-informed reconstruction based on the Raychaudhuri equation ~\citep{Raychaudhuri:1953yv,Ehlers:1961xww}, which provides a general kinematic prior that suppresses unphysical fluctuations in derivative-based reconstructions while preserving their model-independent nature. Using this framework, we identify a localized deviation from the Planck-$\Lambda$CDM prediction for $H(z)$ at $z\simeq0.47$, preferred by current low-redshift distance measurements.

The structure of this paper is as follows. In Section~\ref{sec2}, we outline the methodology for reconstructing cosmological distances directly from the data and present the physics-informed reconstruction approach based on the Raychaudhuri equation. Section~\ref{data} summarizes the observational datasets used in this work. Section~\ref{ra_reira} compares the purely data-driven and physics-informed reconstruction approaches and quantifies the impact of the Raychaudhuri prior. Section~\ref{ten_plc} presents the reconstructed expansion history and the evidence for localized anomalies in the late-time expansion history. Finally, Section~\ref{conclu} summarizes our main results and presents our concluding remarks.

\section{Reconstruction of Cosmological Distances}\label{sec2}

Cosmological distances inferred from redshifts, angular sizes, and fluxes are key geometric observables in modern cosmology and play a central role in probing the expansion history of the Universe. In a spatially flat Friedmann--Lema\^{i}tre--Robertson--Walker (FLRW) spacetime,
\begin{equation}
    \mathrm{d}s^2 = -c^2\mathrm{d}t^2 + a^2(t)\,\big[\mathrm{d}r^2 + r^2\,\mathrm{d}\Omega^2\big],
\end{equation}
these distances are completely determined by the expansion history through the Hubble parameter $H=\dot a(t)/a(t)$, where $a(t)$ is the scale factor. Consequently, cosmological distance measurements provide one of the most direct probes of the expansion history and offer a powerful avenue for investigating observational tensions, anomalies, and possible departures from the standard cosmological model.

There are three principal distance measures directly connected to observations~\citep{Hogg:1999ad}: the comoving distance $d_M(z)$, the angular-diameter distance $d_A(z)$, and the luminosity distance $d_L(z)$, defined as
\begin{equation}\label{disrel}
\begin{split}
    d_M(z) &\equiv c \int_{t}^{t_0} \frac{\mathrm{d}t}{a(t)}
    = c \int_0^z \frac{\mathrm{d}z}{H(z)}, \\
    d_A(z) &= \frac{d_M(z)}{1+z}, \\
    d_L(z) &= (1+z)\, d_M(z),
\end{split}
\end{equation}
where $c$ is the speed of light, and the redshift ($z$) is related to the scale factor through $a=(1+z)^{-1}$.

In this work, we adopt the spline interpolation method introduced in~\citet{Mukherjee:2025ytj} to reconstruct the dimensionless comoving distance,
\begin{equation}
D_M(z)\equiv\frac{H_0\,d_M(z)}{c},
\end{equation}
where $H_0$ is the Hubble constant today. The corresponding dimensionless Hubble distance is then defined as
\begin{equation}
D_H(z)\equiv D_M'(z)=\frac{H_0}{H(z)},
\end{equation}
which is simply the inverse dimensionless expansion rate and is therefore directly related to the expansion history. Here, a prime denotes differentiation with respect to redshift $z$.

\subsection{Reconstruction Methodology}

We reconstruct the quantity of interest using a spline interpolation defined over a set of fixed nodes (``knots''). Let $f(x)$ denote the function to be reconstructed, represented by a spline of order $k$ defined over $n$ knot locations $\{x_1,x_2,\ldots,x_n\}$ with corresponding amplitudes $\{f_1,f_2,\ldots,f_n\}$. The spline consists of piecewise polynomials of degree $k-1$, with continuity enforced up to the $(k-2)$-th derivative at each knot. The knot locations are fixed, while the amplitudes $\{f_m\}$ are treated as free parameters and inferred by fitting the observational data, whose uncertainties are described by the covariance matrix $\mathcal{C}$.

Assuming Gaussian-distributed errors, the log-likelihood is
\begin{equation}
\ln \mathcal{L} =
-\frac{1}{2}\mathbf{r}^{\rm T}\mathcal{C}^{-1}\mathbf{r}
-\frac{1}{2}\ln|\mathcal{C}|
-\frac{N}{2}\ln(2\pi),
\end{equation}
where
\begin{equation}
\mathbf{r}=y_{\rm obs}-f(x_{\rm obs};\{x_m,f_m\})
\end{equation}
denotes the residual vector between the observed data and the spline reconstruction evaluated at the corresponding redshifts.

Bayesian parameter inference is performed using Markov Chain Monte Carlo (MCMC) sampling to explore the posterior distribution of the spline amplitudes. Since spline derivatives are analytic, both the reconstructed function and its derivatives are obtained directly from each posterior sample, allowing a consistent propagation of uncertainties. Similar free-form and knot-based reconstruction techniques have been widely applied in cosmology~\citep{Brumback01091998,Sealfon:2005em,Sahni:2006pa,Planck:2015sxf,Zhao:2017cud,Ormondroyd:2025exu}.

Knot-based splines provide a flexible and largely model-independent framework for reconstructing smooth functions and their derivatives from discrete observations.

We reconstruct the dimensionless comoving distance $D_M(z)$ using a spline of order $k=5$. Following~\citet{Mukherjee:2025ytj}, we place knots at the five characteristic redshifts
\begin{equation}
z=[0.350,\;0.512,\;0.782,\;1.236,\;1.626],
\end{equation}
which were identified as representative locations in the expansion history. We denote these redshifts as $z_1$--$z_5$. In addition, a boundary knot is placed at the highest-redshift data point, $z_6=2.33$. The reconstruction is anchored at the present epoch by imposing the exact boundary conditions
\begin{equation}
D_M(0)=0,
\end{equation}
and
\begin{equation}
D_M'(0)=1,
\end{equation}
which follow directly from the definition of the dimensionless comoving distance. The choice $k=5$ provides sufficient smoothness for the second derivative $D_M''(z)$, which plays an important role in the Raychaudhuri equation-based analysis presented below. As demonstrated in Sec.~\ref{ten_plc}, the reconstruction is stable against variations in both the spline order and the knot configuration.

Rather than reconstructing $D_M(z)$ directly, we interpolate its residuals with respect to the Planck 2018 baseline $\Lambda$CDM model, $D_M^{\rm P18}$. The free parameters therefore consist of the six spline amplitudes
\begin{equation}
\Delta D_M(z_i)\equiv D_M(z_i)-D_M^{\rm P18}(z_i),
\end{equation}
one at each knot location, together with the dimensionless Hubble parameter
\begin{equation}
h\equiv\frac{H_0}{100\,{\rm km\,s^{-1}\,Mpc^{-1}}},
\end{equation}
and the sound horizon at the drag epoch, $r_d$.

We adopt flat priors
\begin{equation}
\Delta D_M(z_i)\in[-0.5,0.5],
\end{equation}
for the spline amplitudes and
\begin{equation}
h\in[0.5,1.2],
\end{equation}
together with the Gaussian prior
\begin{equation}
r_d=147.09\pm0.27~{\rm Mpc},
\end{equation}
consistent with the Planck 2018 $\Lambda$CDM determination. This choice reflects our assumption that the pre-recombination physics remains unchanged, so that the sound horizon at the drag epoch is not modified by new early-Universe physics. In Sec.~\ref{ten_plc}, we investigate the robustness of our results by relaxing this assumption. 

The posterior samples of the spline amplitudes are used to reconstruct $D_M(z)$ over the full redshift range probed by the observations. Since spline derivatives can be evaluated analytically, the corresponding posterior distributions of $D_M'(z)$ and $D_M''(z)$ are obtained directly from the reconstructed splines. The corresponding dimensionful quantities, including the comoving distance $d_M(z)$ and the Hubble parameter $H(z)$, are then recovered by combining the reconstructed dimensionless functions with the posterior distribution of $h$. Throughout this work, we refer to this purely data-driven framework as the \textbf{Reconstruction Algorithm (RA)}.

\subsection{Physics-Informed REIRA}

While the reconstruction described above is entirely data-driven, derivative-based reconstructions are inherently more susceptible to statistical fluctuations than the reconstructed distance itself. It is therefore desirable to incorporate general physical information that stabilizes the reconstruction without introducing assumptions about the nature of dark energy or a specific cosmological model. We achieve this by exploiting the Raychaudhuri equation (RE), a fundamental geometric relation governing the evolution of congruences in spacetime.

For a timelike congruence with four-velocity $u^\mu$ ($u_\mu u^\mu=-1$), the RE is given by~\citep{Raychaudhuri:1953yv,Ehlers:1961xww}
\begin{equation}
\frac{\mathrm{d}\theta}{\mathrm{d}\tau}
=
-\frac{1}{3}\theta^2
-\sigma_{\mu\nu}\sigma^{\mu\nu}
+\omega_{\mu\nu}\omega^{\mu\nu}
+\nabla_\mu A^\mu
-R_{\mu\nu}u^\mu u^\nu,
\end{equation}
where $\theta$ is the expansion scalar, $\tau$ is the proper time, $\sigma_{\mu\nu}$ and $\omega_{\mu\nu}$ are the shear and vorticity tensors, respectively, $A^\mu$ is the four-acceleration, and $R_{\mu\nu}$ is the Ricci tensor.

For a spatially flat FLRW spacetime within General Relativity, this relation reduces to
\begin{equation}
\frac{2}{3}(1+z)HH' - H^2
=
\frac{8\pi G}{3c^2}p(z),
\end{equation}
where $p(z)$ denotes the total cosmic pressure.

At late times, the total cosmic pressure is dominated by the dark energy component, since the matter sector is effectively pressureless and the radiation contribution is negligible. Consequently, it is well motivated to require that $p<0$ over the redshift range probed by the observations considered in this work. Expressed in terms of the reconstructed quantity $D_M$, this condition becomes
\begin{equation}
\label{eq:RE_constraint}
\frac{2}{3}(1+z)\frac{D_M''}{(D_M')^3}
+
\frac{1}{(D_M')^2}
=
-\frac{p}{\rho_{c0}c^2}>0,
\end{equation}
where
\begin{equation}
\rho_{c0}\equiv\frac{3H_0^2}{8\pi G},
\end{equation}
denotes the present-day critical energy density of the Universe.

We impose this inequality as a physical prior over the redshift range $0<z<2.33$. The resulting constraint is remarkably general. It does not depend on the detailed nature of the dark energy sector, whether dark energy interacts with dark matter, or whether the underlying dynamics are canonical or non-canonical. No assumptions are made regarding the separate conservation of the individual cosmic components. The only fundamental assumption is that gravity is described by General Relativity and the Universe is spatially flat.

The Raychaudhuri prior therefore encodes the physically motivated requirement of negative total cosmic pressure in the late-time Universe, translating it into a regularizing constraint on the derivatives of the reconstructed distances, which are the quantities most susceptible to statistical fluctuations.

We refer to this physics-informed framework as the \textbf{Raychaudhuri Equation-Informed Reconstruction Algorithm (REIRA)}.

\begin{figure*}
    \centering
    \includegraphics[width=0.7\linewidth]{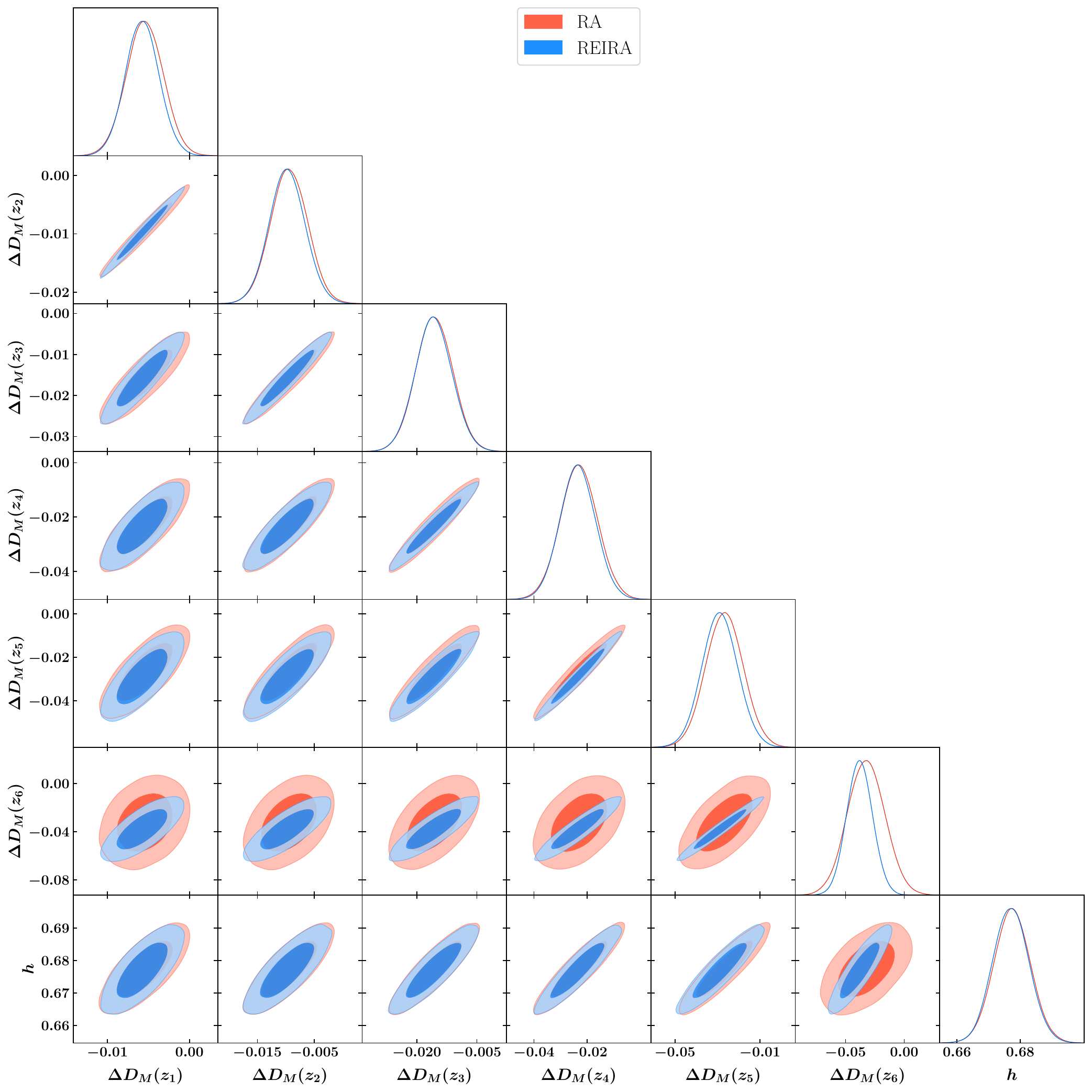}\\[2ex]
    \includegraphics[width=0.245\textwidth]{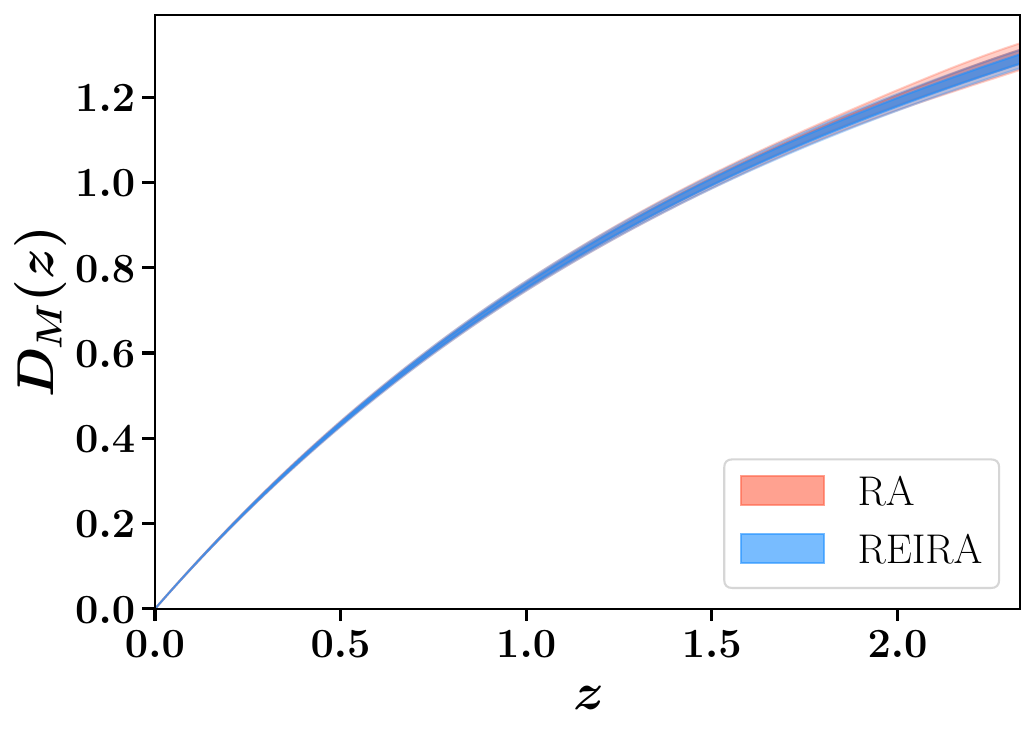}
    \includegraphics[width=0.245\textwidth]{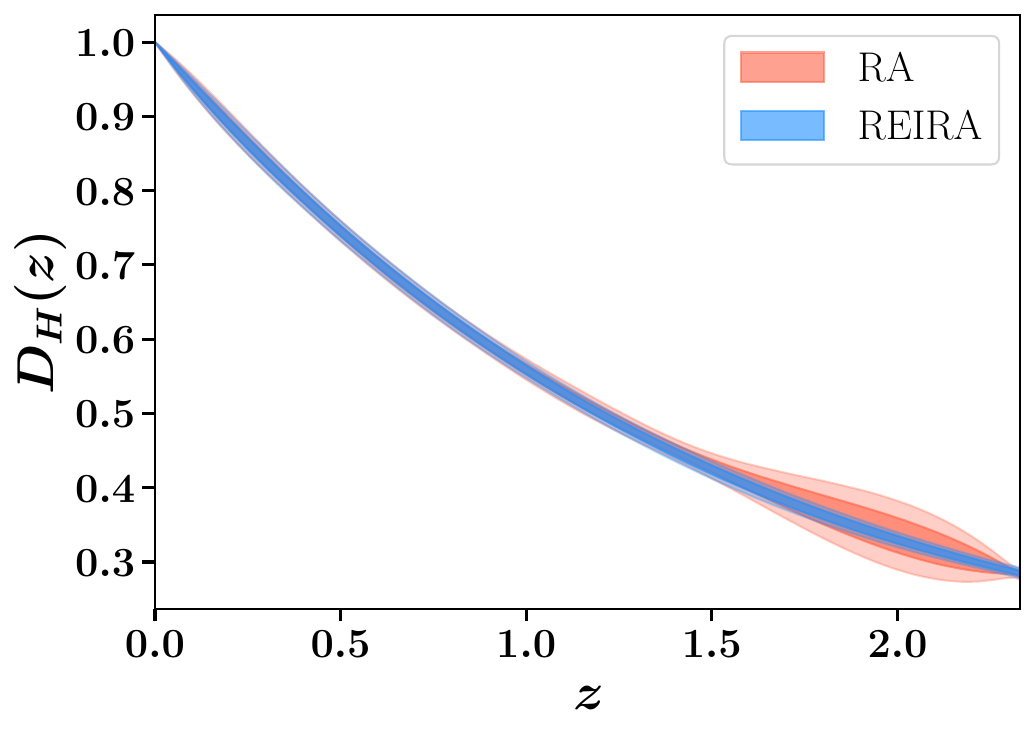}
    \includegraphics[width=0.245\textwidth]{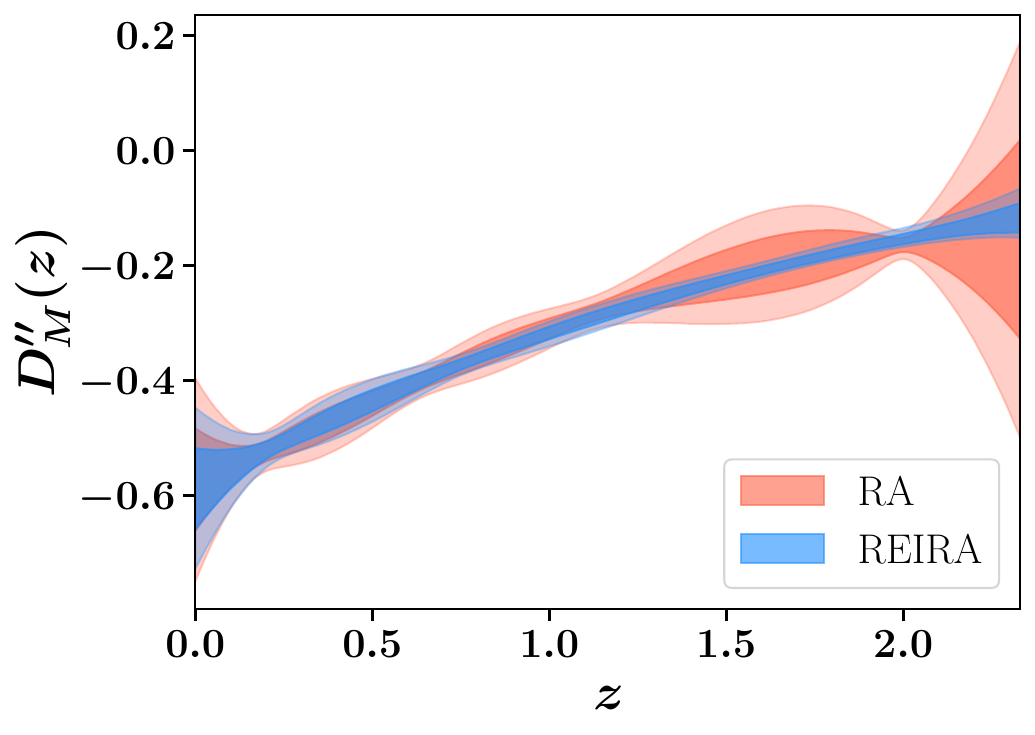}
        \includegraphics[width=0.245\textwidth]{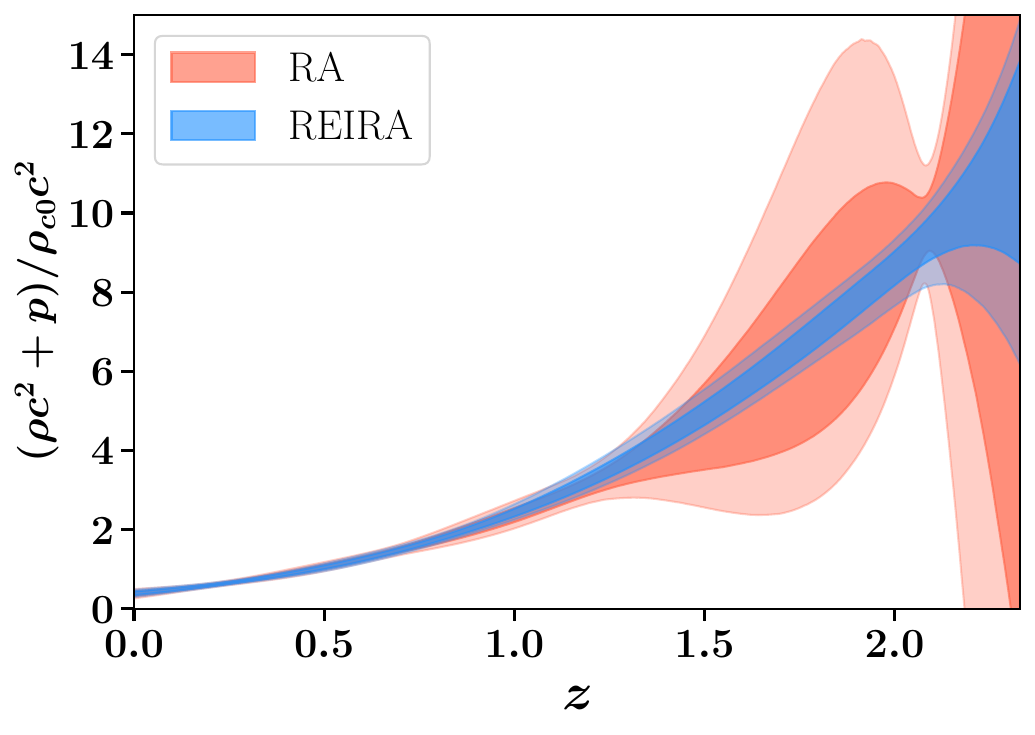}
    \caption{Upper panel: One-dimensional marginalized posterior distributions and two-dimensional 68\% and 95\% confidence contours for $\Delta D_M(z_i)$ and $h$ obtained with RA and REIRA. Lower panel: Reconstructed $D_M(z)$, $D_H(z)$, $D_M^{\prime\prime}(z)$, and $\frac{\rho(z)c^2+p(z)}{\rho_{c0}c^2}$ for the two reconstruction methods. Darker and lighter shaded regions denote the 68\% and 95\% confidence intervals, respectively.}
     \label{dis_evol}
\end{figure*}

\section{Observational Datasets}\label{data}

In this work, we use the following observational datasets:

\begin{itemize}

\item \textbf{Baryon Acoustic Oscillations (BAO):}
We use the 13 correlated BAO distance measurements from the DESI Data Release 2 (DESI-DR2), spanning the redshift range $0.1<z<4.2$. These measurements, compiled in Table~IV of~\citet{DESI:2025zgx}, provide constraints on the expansion history over a broad range of cosmic time. Throughout this paper, we refer to this dataset as \textit{DESI-DR2}.

\item \textbf{Type Ia Supernovae (SNIa):}
We use the DES-Dovekie compilation from the five-year Dark Energy Survey Supernova program~\citep{DES:2025sig}. This dataset provides luminosity-distance measurements over the low- to intermediate-redshift Universe and is referred to throughout this paper simply as \textit{DES}.

\end{itemize}

\section{RA vs REIRA}\label{ra_reira}

We reconstruct the key cosmological observables using both RA and REIRA for the DESI-DR2 + DES dataset combination. Both reconstruction methods have been extensively validated using mock datasets, as discussed in Appendix~\ref{mock}.

The upper panel of Fig.~\ref{dis_evol} compares the posterior distributions of the spline amplitudes, $\Delta D_M(z_i)$, obtained with RA and REIRA. The inclusion of the Raychaudhuri prior significantly tightens the constraints on the reconstructed distances, with the improvement becoming progressively more pronounced towards higher redshifts, where the observational coverage is relatively sparse. A smaller, but still noticeable, improvement is also observed at lower redshifts.

The lower panel of Fig.~\ref{dis_evol} compares the reconstructed evolution of $D_M(z)$, $D_H(z)$, and $D_M''(z)$. REIRA consistently yields tighter constraints than the purely data-driven reconstruction, with the largest improvement occurring for quantities involving higher-order derivatives. This behaviour is expected because the Raychaudhuri prior directly constrains combinations of $D_M'(z)$ and $D_M''(z)$, thereby suppressing unphysical fluctuations in the reconstructed distance.

The improvement is particularly evident in the reconstruction of $D_H(z)$, shown in the second panel of the lower row of Fig.~\ref{dis_evol}. In the RA reconstruction, the uncertainty grows rapidly beyond $z\sim1.5$ owing to the limited observational information available between the highest-redshift DES supernova and the DESI measurement at $z=2.33$. By suppressing unphysical spline excursions in this sparsely constrained region, the Raychaudhuri prior produces a smoother and more stable reconstruction of the high-redshift expansion history. An even larger reduction in the uncertainty is observed for $D_M''(z)$, which depends explicitly on the second derivative of the reconstructed distance. Importantly, while REIRA substantially reduces the uncertainties, it does not qualitatively alter the reconstructed expansion history relative to RA. Instead, it suppresses poorly constrained fluctuations that are not supported by the data, thereby yielding a more stable reconstruction.

To examine the Null Energy Condition (NEC), $R_{\mu\nu}k^{\mu}k^{\nu}\geq0$, where $k^{\mu}$ is a null four-vector, we also reconstruct the dimensionless combination of the total energy density and pressure,
\begin{equation}
\frac{\rho(z)c^2+p(z)}{\rho_{c0}c^2},
\end{equation}
shown in the bottom panel of Fig.~\ref{dis_evol}. We find no evidence for a violation of the NEC, $\rho c^2+p\ge0$, across the entire redshift range probed by the data in either the RA or REIRA reconstruction. The expansion history inferred from the DESI and DES observations is therefore fully consistent with the NEC.

This result is particularly relevant in light of recent indications of a phantom-divide crossing in reconstructions of the dark energy equation of state~\citep{DESI:2025fii,Ozulker:2025ehg}. Since the NEC depends on the total cosmic fluid rather than the dark energy component alone, our results suggest that an apparent phantom behaviour inferred from an effective dark energy equation of state does not necessarily imply a violation of the fundamental energy conditions. Instead, it may reflect the effective description of a more complex dark sector~\citep{Caldwell:2025inn,Liu:2025bss}.

Compared with the purely data-driven RA reconstruction, REIRA substantially suppresses the poorly constrained fluctuations that arise at high redshift owing to the limited constraining power of the data. At the same time, it preserves the overall reconstruction while significantly reducing the associated uncertainties, without introducing a specific dark energy model. Motivated by these advantages, we adopt REIRA as the default reconstruction framework throughout the remainder of this work unless stated otherwise.

\begin{figure*}
\centering
\includegraphics[width=0.45\linewidth]{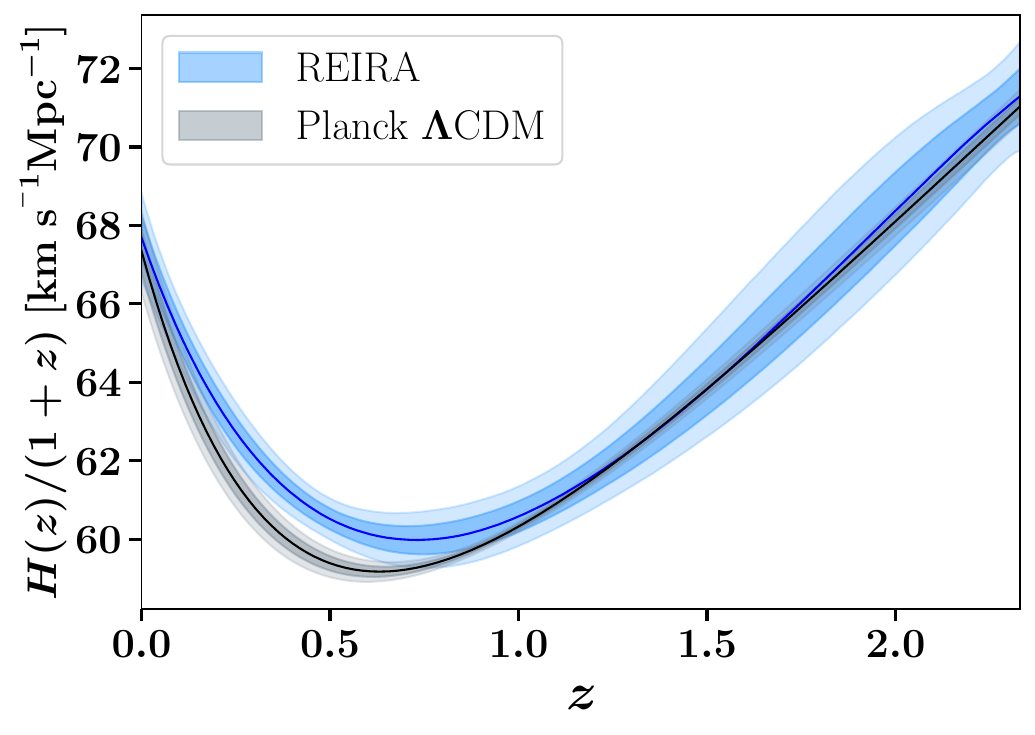}
~~\includegraphics[width=0.45\linewidth]{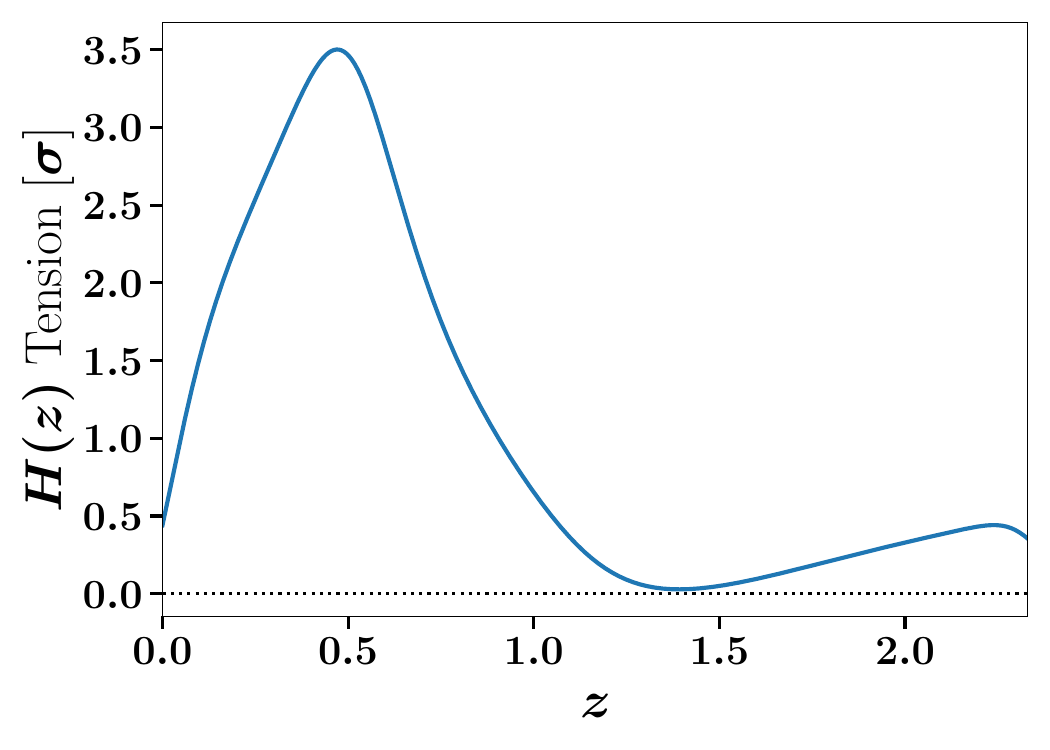}\\
\includegraphics[width=0.45\linewidth]{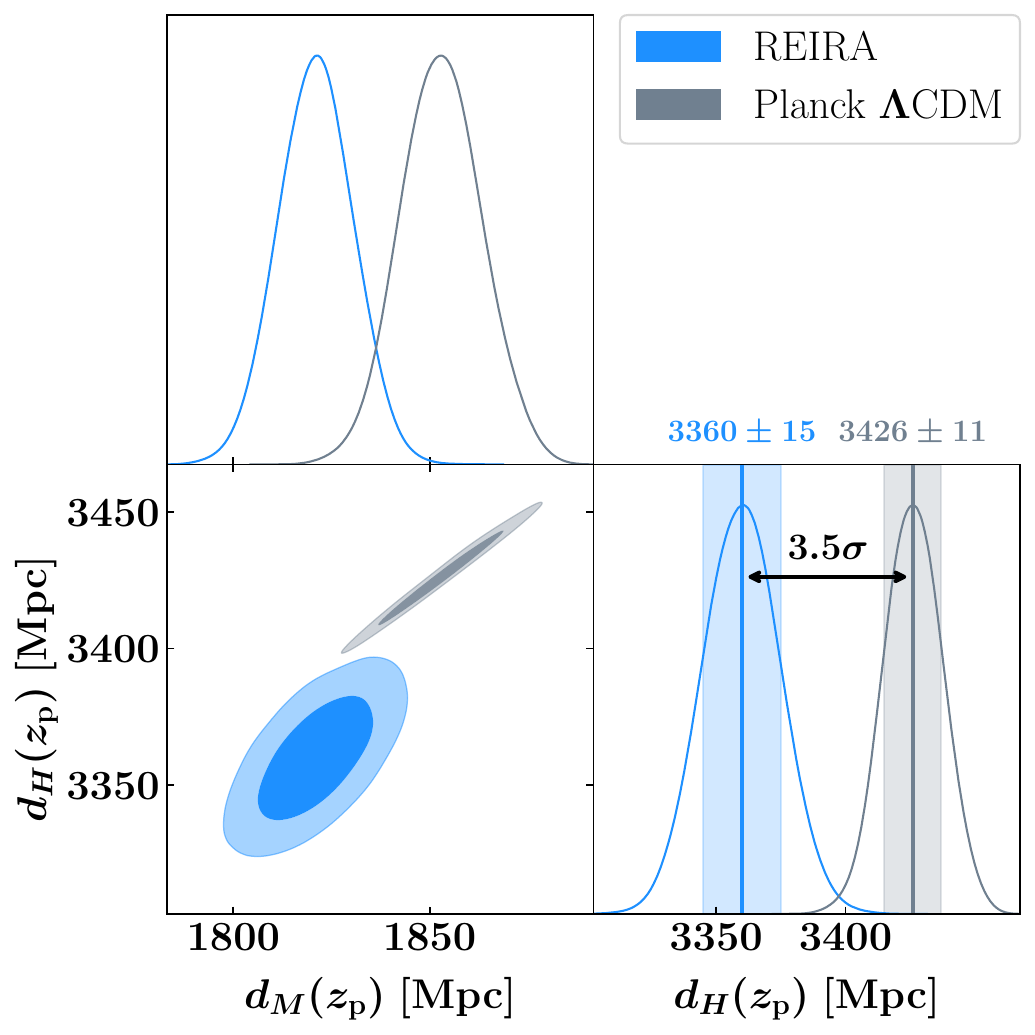}
\caption{Upper panel: Evolution of $H(z)/(1+z)$ reconstructed with REIRA compared with the Planck 2018 $\Lambda$CDM prediction (left), and the corresponding tension with respect to the Planck 2018 $\Lambda$CDM model as a function of redshift (right). Darker and lighter shaded regions denote the 68\% and 95\% confidence intervals, respectively, while the solid curves show the best-fit evolution. Lower panel: One-dimensional marginalized posterior distributions and two-dimensional 68\% and 95\% confidence contours for $d_M$ and $d_H$ (in Mpc) at $z_{\rm p}=0.47$.}
\label{hubz}
\end{figure*}

\section{Localized Deviation from Planck $\Lambda$CDM H(z) Prediction}\label{ten_plc}

Having established the advantages of REIRA over RA, we now investigate the extent to which the reconstructed expansion history differs from the Planck 2018 $\Lambda$CDM prediction. The upper-left panel of Fig.~\ref{hubz} compares the reconstructed evolution of $H(z)/(1+z)$ with the corresponding Planck 2018 $\Lambda$CDM prediction, while the upper-right panel shows the redshift dependence of the tension relative to Planck. To quantify this discrepancy, we define the tension statistic as
\begin{equation}
T=
\frac{\mu_{\rm REIRA}-\mu_{\rm P18}}
{\sqrt{\sigma_{\rm REIRA}^2+\sigma_{\rm P18}^2}},
\end{equation}
where $\mu_{\rm REIRA}$ and $\sigma_{\rm REIRA}$ denote the posterior mean and $1\sigma$ uncertainty obtained from the REIRA reconstruction, while $\mu_{\rm P18}$ and $\sigma_{\rm P18}$ are the corresponding quantities predicted by the Planck 2018 $\Lambda$CDM model.

A prominent feature of the reconstructed expansion history is a localized deviation from the Planck prediction over the redshift range $0.3\lesssim z\lesssim0.6$. The tension reaches its maximum at $z\simeq0.47$, where the reconstructed Hubble parameter differs from the Planck $\Lambda$CDM prediction by approximately $3.5\sigma$. To further illustrate this discrepancy, the lower panel of Fig.~\ref{hubz} shows the one-dimensional marginalized posterior distributions together with the two-dimensional 68\% and 95\% confidence contours for $d_M$ and $d_H$ at $z_{\rm p}=0.47$. The figure demonstrates that the discrepancy is driven primarily by the reconstructed expansion rate rather than by the distance itself.

For completeness, the REIRA reconstruction yields
\begin{equation}
H_0=67.71\pm0.57~{\rm km\,s^{-1}\,Mpc^{-1}},
\end{equation}
fully consistent with the Planck 2018 $\Lambda$CDM value of
\begin{equation}
H_0=67.37\pm0.54~{\rm km\,s^{-1}\,Mpc^{-1}},
\end{equation}
reported in~\citet{Planck:2018vyg}. This agreement is expected because the sound horizon is constrained by a Planck-based prior. Consequently, the localized anomaly identified here cannot be attributed to a mismatch in the inferred value of $H_0$, but instead reflects a genuine feature in the reconstructed late-time expansion history.

To assess the robustness of the localized anomaly, we repeat the reconstruction under a variety of alternative assumptions and dataset combinations. Unless otherwise stated, these tests are performed using the purely data-driven RA framework to demonstrate that the observed feature is not induced by the Raychaudhuri prior.

\begin{figure*}
\centering

\begin{subfigure}{0.33\textwidth}
\includegraphics[width=\linewidth]{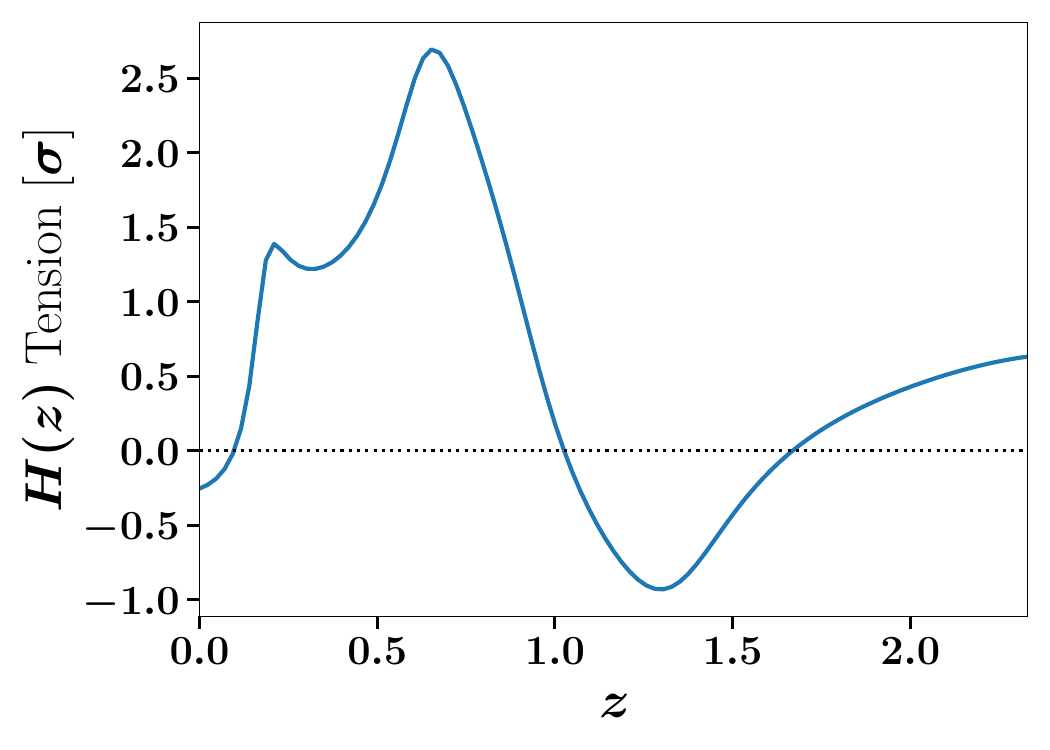}
\caption{}\label{ten1}
\end{subfigure}
\begin{subfigure}{0.33\textwidth}
\includegraphics[width=\linewidth]{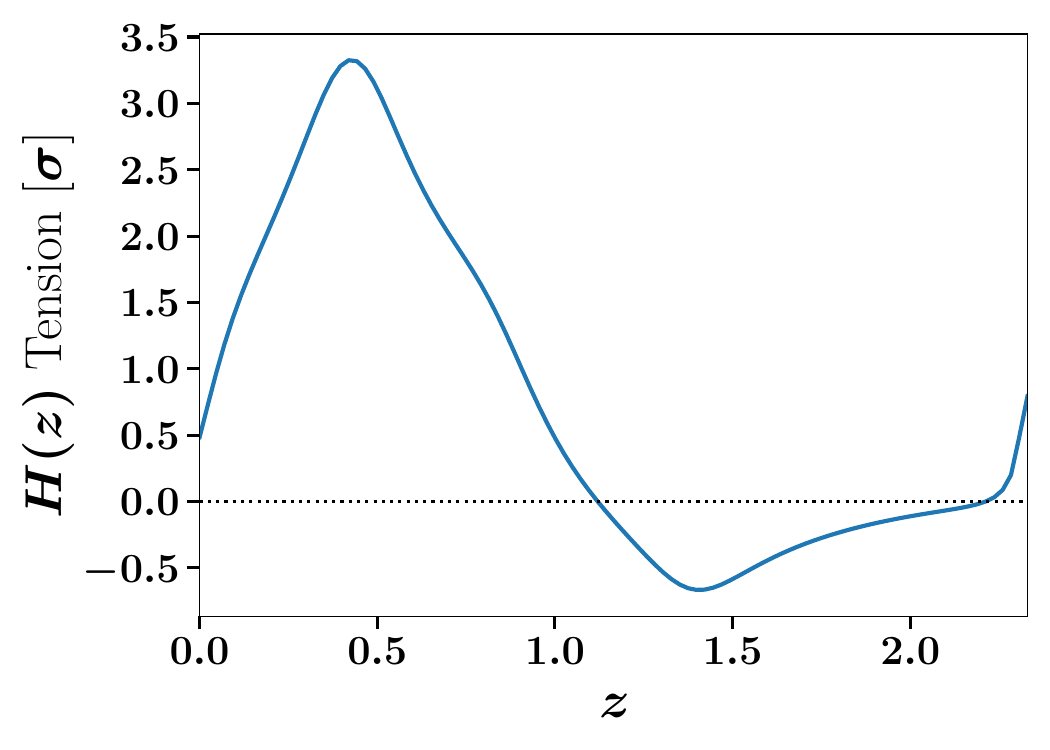}
\caption{}\label{ten2}
\end{subfigure}
\begin{subfigure}{0.33\textwidth}
\includegraphics[width=\linewidth]{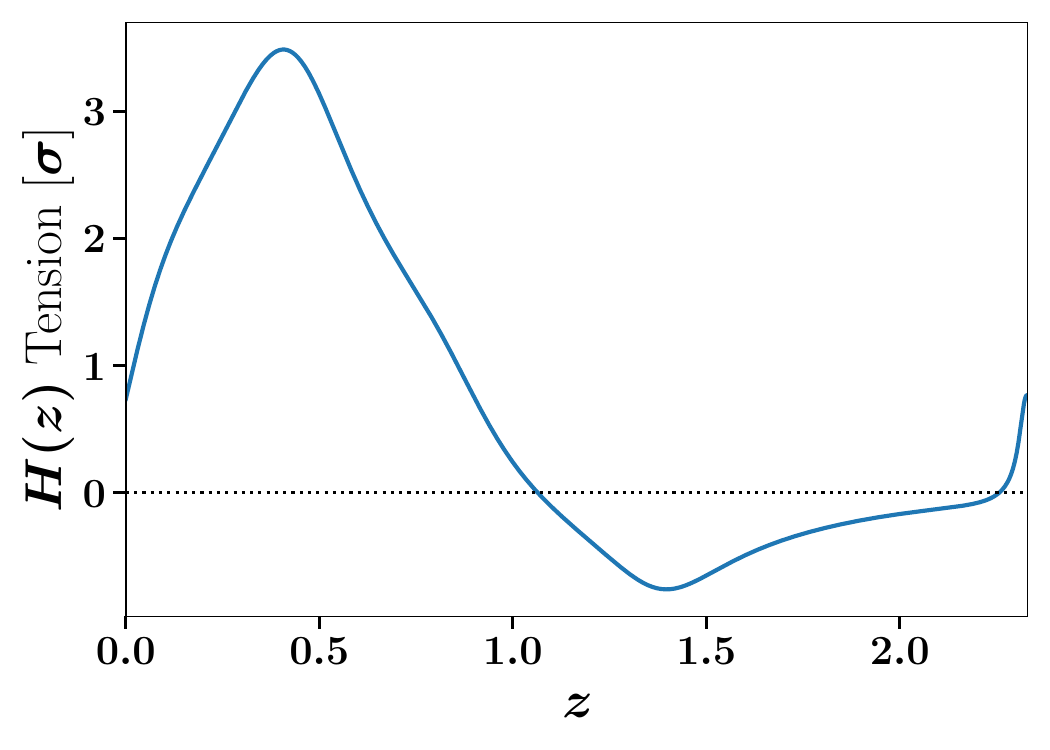}
\caption{}\label{ten3}
\end{subfigure}

\begin{subfigure}{0.33\textwidth}
\includegraphics[width=\linewidth]{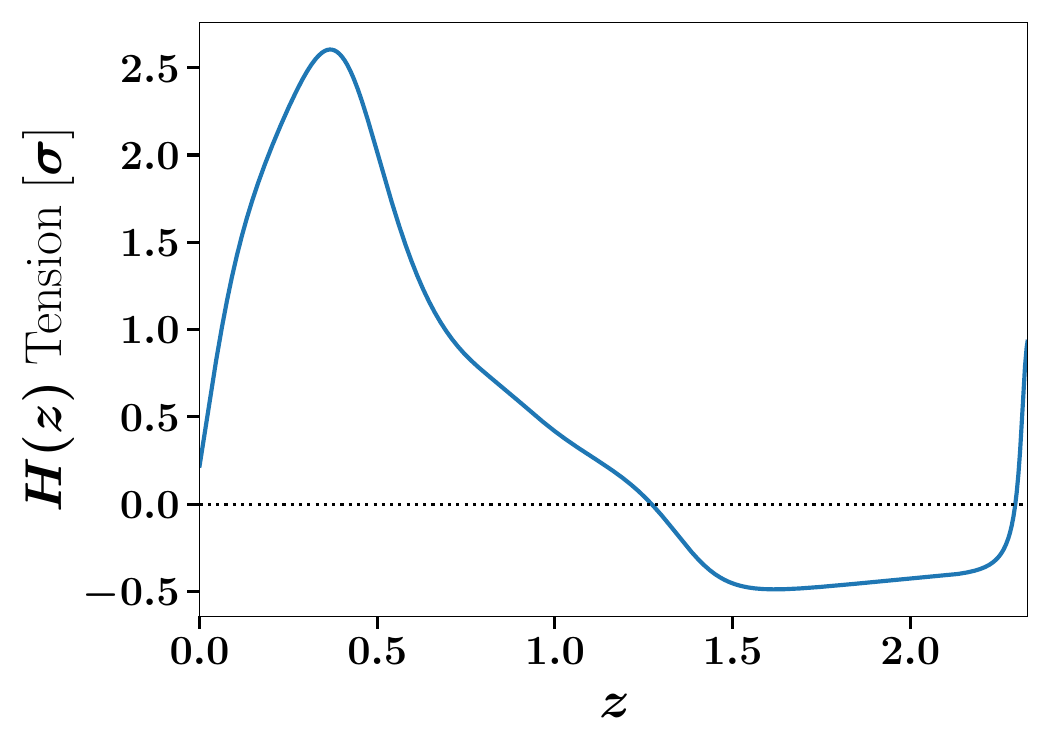}
\caption{}\label{ten4}
\end{subfigure}
\begin{subfigure}{0.33\textwidth}
\includegraphics[width=\linewidth]{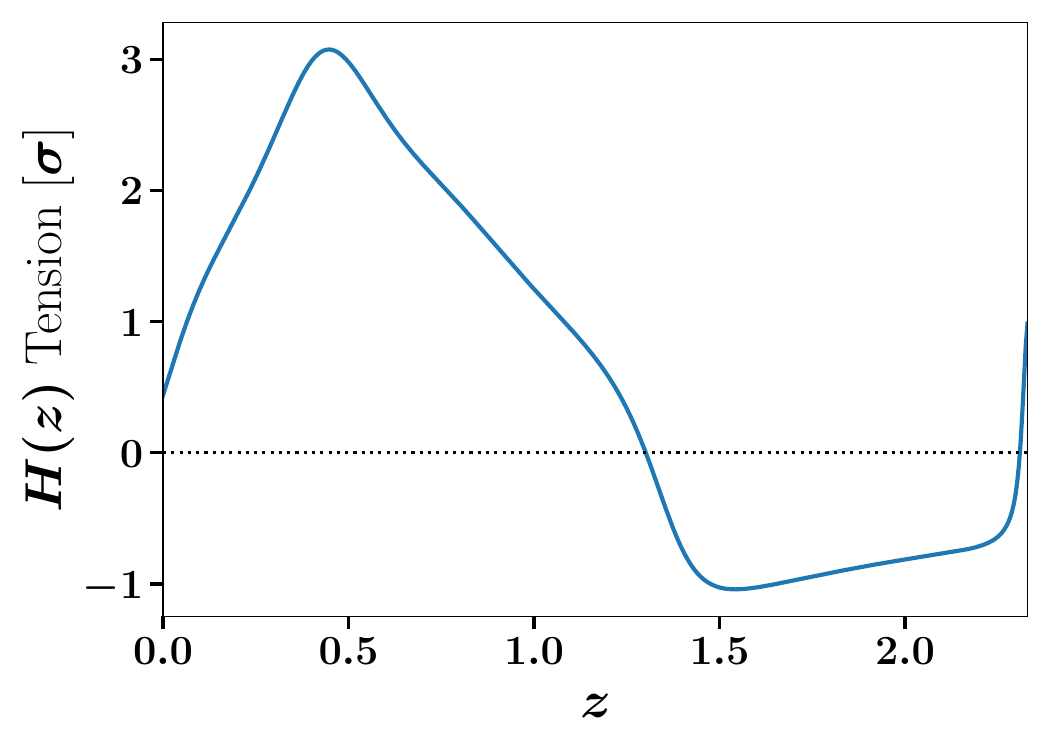}
\caption{}\label{ten5}
\end{subfigure}
\begin{subfigure}{0.33\textwidth}
\includegraphics[width=\linewidth]{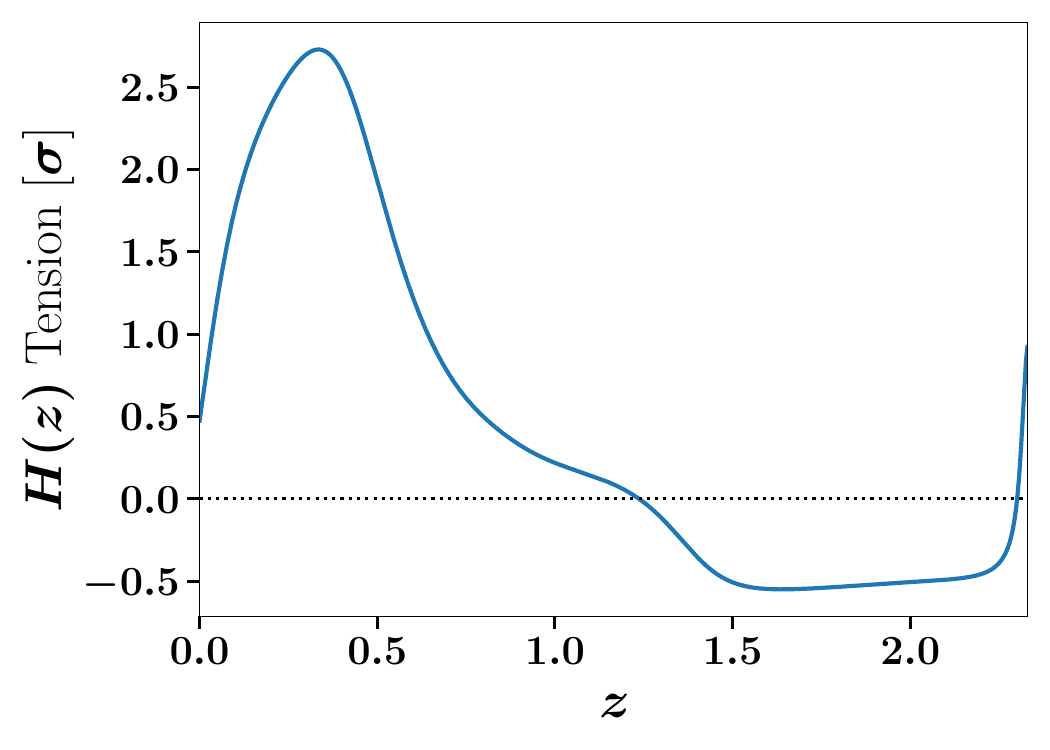}
\caption{}\label{ten6}
\end{subfigure}

\begin{subfigure}{0.33\textwidth}
\includegraphics[width=\linewidth]{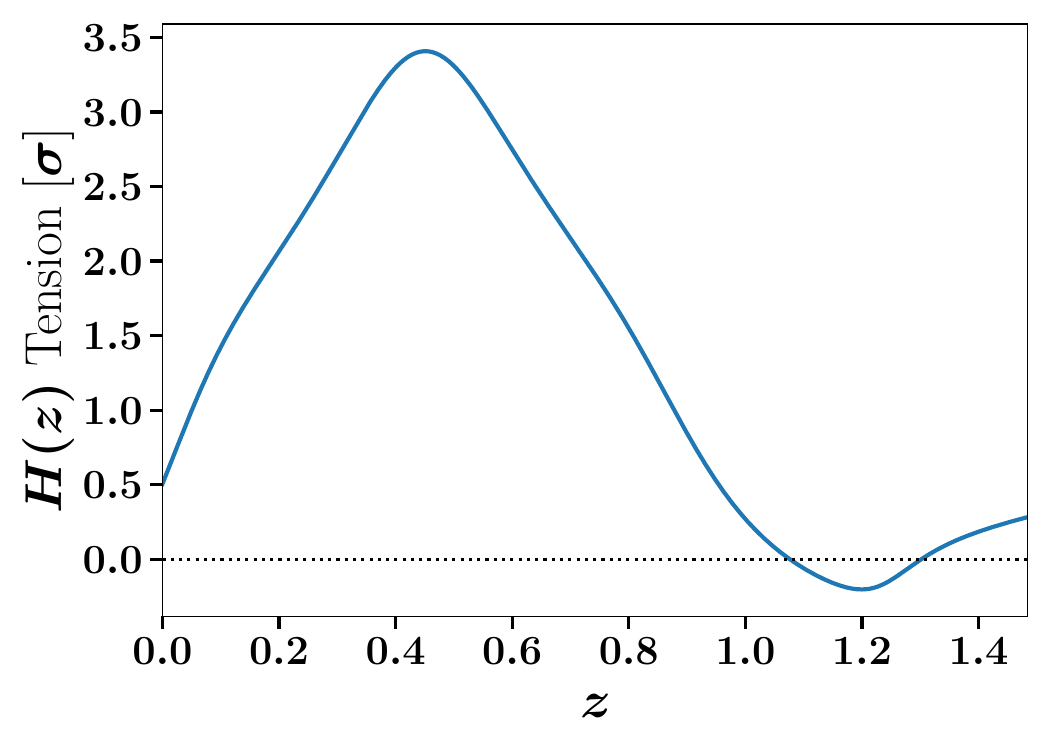}
\caption{}\label{ten7}
\end{subfigure}
\begin{subfigure}{0.33\textwidth}
\includegraphics[width=\linewidth]{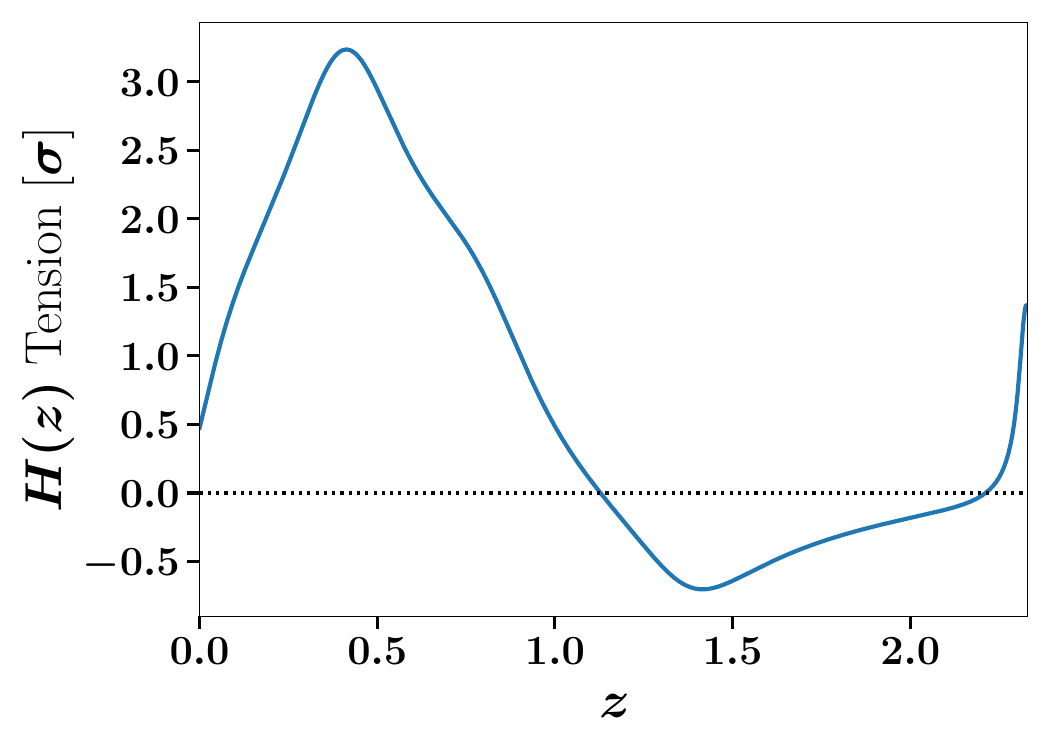}
\caption{}\label{ten8}
\end{subfigure}
\begin{subfigure}{0.33\textwidth}
\includegraphics[width=\linewidth]{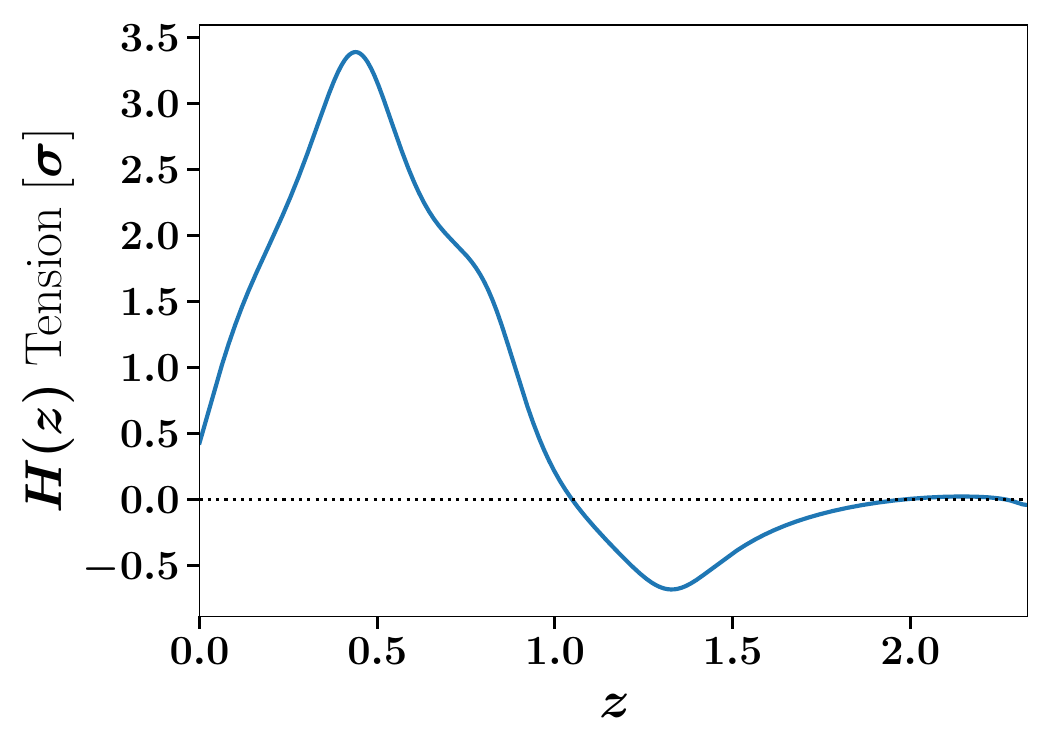}
\caption{}\label{ten9}
\end{subfigure}
\caption{
Robustness tests of the reconstructed tension profile. Panels (a)--(i) show the results obtained using: (a) DESI-DR2 only; (b) DESI-DR2+DES; (c) DESI-DR2+DES without {\bf \texttt{LRG1}}; (d) DESI-DR2+DES without {\bf \texttt{LRG2}}; (e) DESI-DR2+DES without {\bf \texttt{LRG3+ELG1}}; (f) DESI-DR2+DES without {\bf \texttt{LRG1}} and {\bf \texttt{LRG2}}; (g) DESI-DR2+DES without the {\bf \texttt{Lya}} measurements; (h) DESI-DR2+DES with the first five knots placed at the effective redshifts of the DESI-DR2 measurements; and (i) DESI-DR2+DES using a spline of order $k=4$. All reconstructions are performed using the RA framework.
}
\label{tensions_diff}
\end{figure*}

We first reconstruct the expansion history using the DESI-DR2 dataset alone. The resulting tension profile, shown in Fig.~\ref{ten1}, already exhibits a pronounced localized excess exceeding $2.5\sigma$. Including the DES Dovekie supernova sample further increases the significance of this feature, producing a tension profile qualitatively similar to that obtained with REIRA and reaching a maximum significance of approximately $3.3\sigma$ at $z\simeq0.47$ (Fig.~\ref{ten2}).

To determine whether the anomaly is driven by a particular DESI tracer, we repeat the reconstruction after removing the {\bf \texttt{LRG1}}, {\bf \texttt{LRG2}}, and {\bf \texttt{LRG3+ELG1}} measurements individually, as well as removing {\bf \texttt{LRG1}} and {\bf \texttt{LRG2}} simultaneously. The resulting tension profiles are shown in Figs.~\ref{ten3}--\ref{ten6}. We also test the impact of the {\bf \texttt{Lya}} measurements by removing both {\bf \texttt{Lya}} data points together with the boundary knot at $z=2.33$, restricting the reconstruction to the highest remaining redshift, $z=1.484$. The corresponding result is shown in Fig.~\ref{ten7}.

Finally, we investigate the dependence of the results on the reconstruction methodology itself. We consider an alternative knot configuration in which the first five knots are placed at the effective redshifts of the DESI-DR2 measurements (Fig.~\ref{ten8}), and also repeat the analysis using a lower-order spline with $k=4$ (Fig.~\ref{ten9}). Although the exact significance varies slightly among the different tests, all reconstructions consistently exhibit a localized excess over the redshift interval $0.3\lesssim z\lesssim0.6$, with the maximum tension occurring near $z\simeq0.47$. The persistence of this feature under substantial changes to both the dataset and the reconstruction methodology strongly suggests that it is not an artifact of any individual measurement or reconstruction choice.

\begin{figure*}
    \centering
    \includegraphics[width=0.45\linewidth]{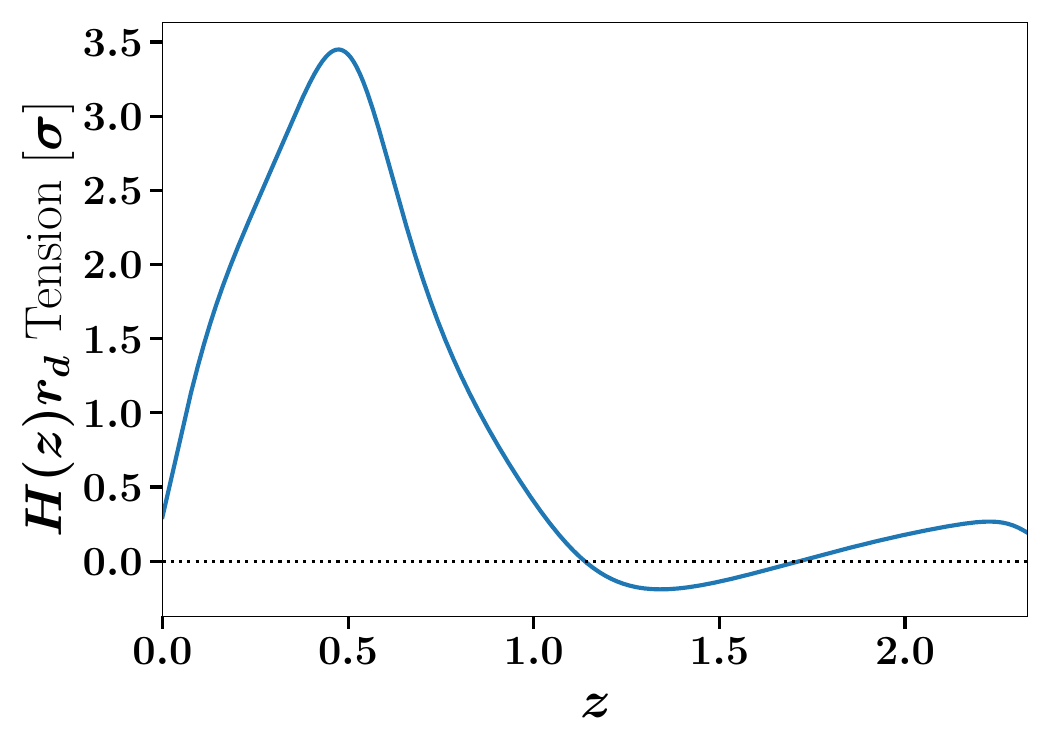} \includegraphics[width=0.45\linewidth]{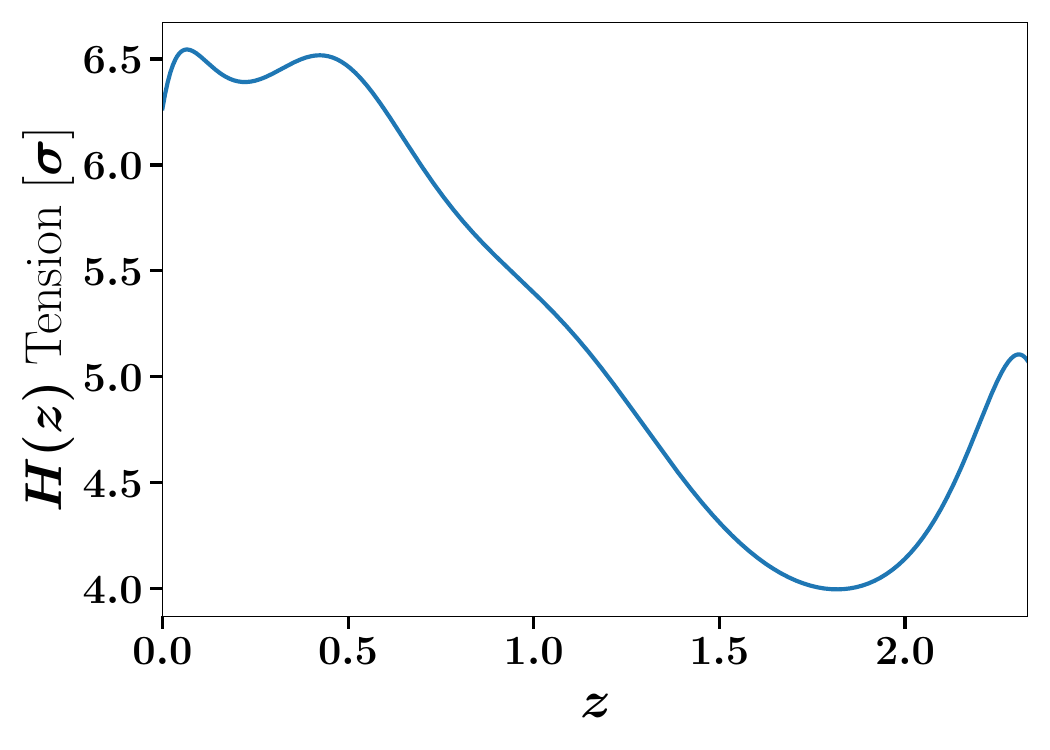}
\caption{
Robustness of the localized low-redshift expansion-rate anomaly under alternative assumptions using REIRA. Left panel: Tension profile obtained by sampling the parameter combination $hr_d$ directly, without imposing an external prior on the sound horizon $r_d$. Right panel: Tension profile obtained by replacing the Planck-based constraint with the H0 Distance Network (H0DN) prior on $H_0$.
}
    \label{tenh0dn}
\end{figure*}

Taken together, these results provide compelling evidence for a localized deviation from the Planck 2018 $\Lambda$CDM prediction, with a maximum significance of approximately $3.5\sigma$ at $z\simeq0.47$. The feature is robust against substantial modifications to both the reconstruction methodology and the dataset. In particular, it is not driven by any individual DESI tracer, making it unlikely that the anomaly originates from an unknown systematic affecting a specific redshift bin.

To further investigate whether the result depends on the adopted calibration of the sound horizon, we perform two additional analyses. First, we repeat the REIRA reconstruction by sampling the parameter combination $hr_d$ directly, thereby removing the external prior on $r_d$. The resulting tension profile for $H(z)r_d$, shown in the left panel of Fig.~\ref{tenh0dn}, retains the localized excess near $z\simeq0.47$. We obtain
\begin{equation}
hr_d=99.62\pm0.84~{\rm Mpc},
\end{equation}
which is fully consistent with the Planck 2018 $\Lambda$CDM value
\begin{equation}
hr_d=99.23\pm0.78~{\rm Mpc},
\end{equation}
reported by~\citet{Planck:2018vyg}. The persistence of the anomaly therefore cannot be attributed to the adopted prior on $r_d$.

As a complementary test, we replace the Planck-based sound-horizon prior with the H0DN prior~\citep{H0DN:2025lyy},
\begin{equation}
H_0=73.50\pm0.81~{\rm km\,s^{-1}\,Mpc^{-1}},
\end{equation}
and repeat the reconstruction. As shown in the right panel of Fig.~\ref{tenh0dn}, the localized excess remains clearly visible.

The persistence of the anomaly under both alternative calibrations demonstrates that it is neither an artifact of the adopted sound-horizon prior nor a consequence of the external $H_0$ calibration. Instead, it appears to be a genuine feature preferred by the low-redshift distance measurements.

\section{Summary and Conclusions}\label{conclu}

In this work, we reconstructed the late-time expansion history of the Universe using the latest DESI-DR2 BAO measurements together with the DES Dovekie SNIa compilation. By combining a model-independent spline reconstruction of cosmological distances with the Raychaudhuri Equation-Informed Reconstruction Algorithm (REIRA), we investigated the consistency of the expansion history inferred from current low-redshift observations with the Planck 2018 $\Lambda$CDM model. Compared with the purely data-driven RA, REIRA provides a more stable reconstruction by suppressing poorly constrained fluctuations in derivative-dependent quantities while preserving the expansion history preferred by the data.

Our analysis reveals a localized deviation from the Planck 2018 $\Lambda$CDM prediction in the reconstructed Hubble expansion rate over the redshift interval $0.3\lesssim z\lesssim0.6$. The tension exceeds $2\sigma$ across this region and reaches a maximum significance of approximately $3.5\sigma$ at $z\simeq0.47$. Importantly, this feature is already present in the purely data-driven RA reconstruction, demonstrating that it is not induced by the Raychaudhuri prior.

We carried out an extensive suite of robustness tests to investigate the origin of this anomaly. The localized excess persists when individual DESI tracers are removed, including the {\bf\texttt{LRG1}}, {\bf\texttt{LRG2}}, {\bf\texttt{LRG3+ELG1}}, and {\bf\texttt{Lya}} samples, as well as when {\bf\texttt{LRG1}} and {\bf\texttt{LRG2}} are simultaneously excluded. The result also remains stable under substantial modifications of the reconstruction methodology, including changes to the spline order and knot configuration. Furthermore, the anomaly is insensitive to the adopted calibration of the sound horizon and to the external $H_0$ prior. Mock analyses based on the Planck 2018 $\Lambda$CDM fiducial model demonstrate unbiased reconstructions with well-calibrated uncertainties, while the observed $3.5\sigma$ deviation lies well outside the distribution obtained from 100 independent mock realizations, resulting an empirical $p < 0.01$ independent of any distributional assumption. Taken together, these tests strongly suggest that the reconstructed feature is not an artifact of the reconstruction methodology, tracer selection, external calibration, or statistical fluctuations.

The persistence of the localized anomaly under both Planck- and H0DN-based calibrations carries potentially important cosmological implications. Since modifications to pre-recombination physics primarily alter the global calibration of the sound horizon, our results suggest that early-Universe physics alone may be insufficient to account for a localized discrepancy in the late-time expansion history. If confirmed, the observed feature could instead point to previously unrecognized late-time physics or to residual observational systematics affecting a narrow redshift interval.

Future observations from DESI, Euclid, LSST, Roman, and other next-generation cosmological surveys will provide substantially more precise measurements of the expansion history and will be crucial for determining whether this localized deviation represents a statistical fluctuation, an unidentified systematic effect, or the first indication of new late-time cosmological physics.

\software{numpy~\citep{Harris:2020xlr},
          scipy~\citep{Virtanen:2019joe}, 
          matplotlib~\citep{Hunter:2007ouj},
          emcee~\citep{Foreman-Mackey:2012any},
          GetDist~\citep{Lewis:2019xzd}.
          }

\begin{acknowledgments}
SGC acknowledges funding from the Anusandhan National Research Foundation (ANRF), Govt. of India, under the National Post-Doctoral Fellowship (File no. PDF/2023/002066). PM acknowledges funding from ANRF, Govt. of India, under the National Post-Doctoral Fellowship (File no. PDF/2023/001986). EDV is supported by a Royal Society Dorothy Hodgkin Research Fellowship. AAS acknowledges the funding from ANRF, Govt. of India, under the research grant no. CRG/2023/003984. We acknowledge the use of the HPC facility, Pegasus, at IUCAA, Pune, India. This article/publication is based upon work from COST Action CA21136- ``Addressing observational tensions in cosmology with systematics and fundamental physics (CosmoVerse)'', supported by COST (European Cooperation in Science and Technology).
\end{acknowledgments}

          \bibliography{biblio}
\bibliographystyle{aasjournalv7}

\appendix

\section{Mock Analysis}\label{mock}

We validate our reconstruction methodology using mock datasets based on the best-fit Planck 2018 $\Lambda$CDM cosmology. Specifically, we generate DESI-DR2- and DES-like mock observations from the fiducial model using the covariance matrices of the corresponding datasets.

\begin{figure}
\centering
\includegraphics[width=0.23\textwidth]{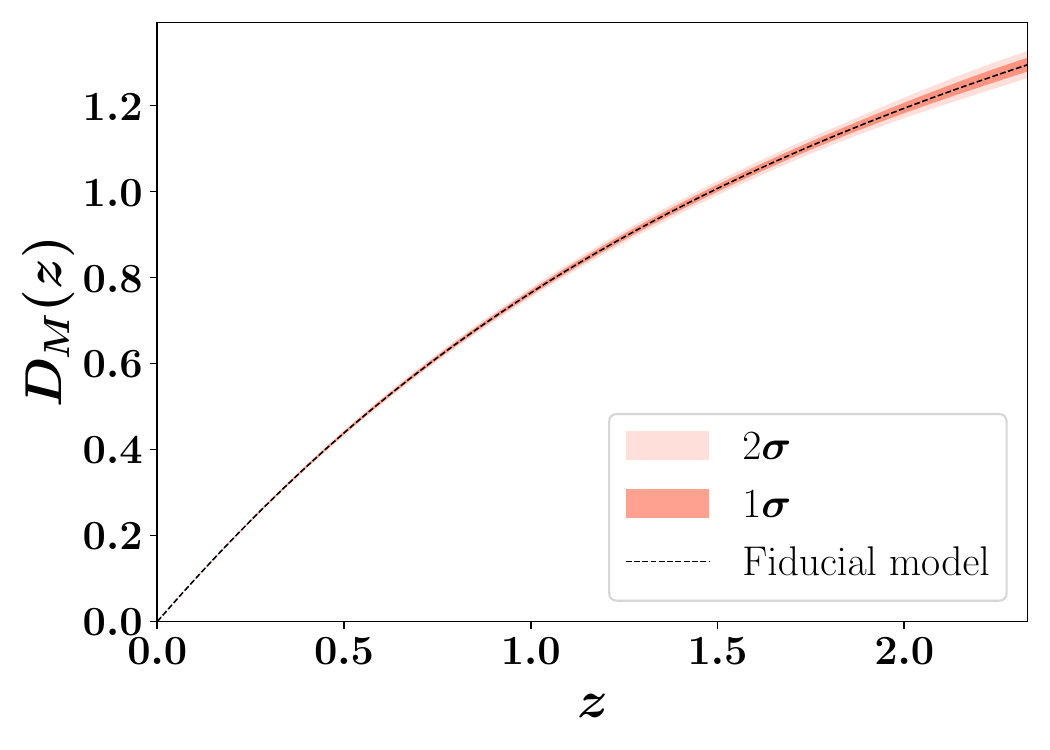}
\includegraphics[width=0.23\textwidth]{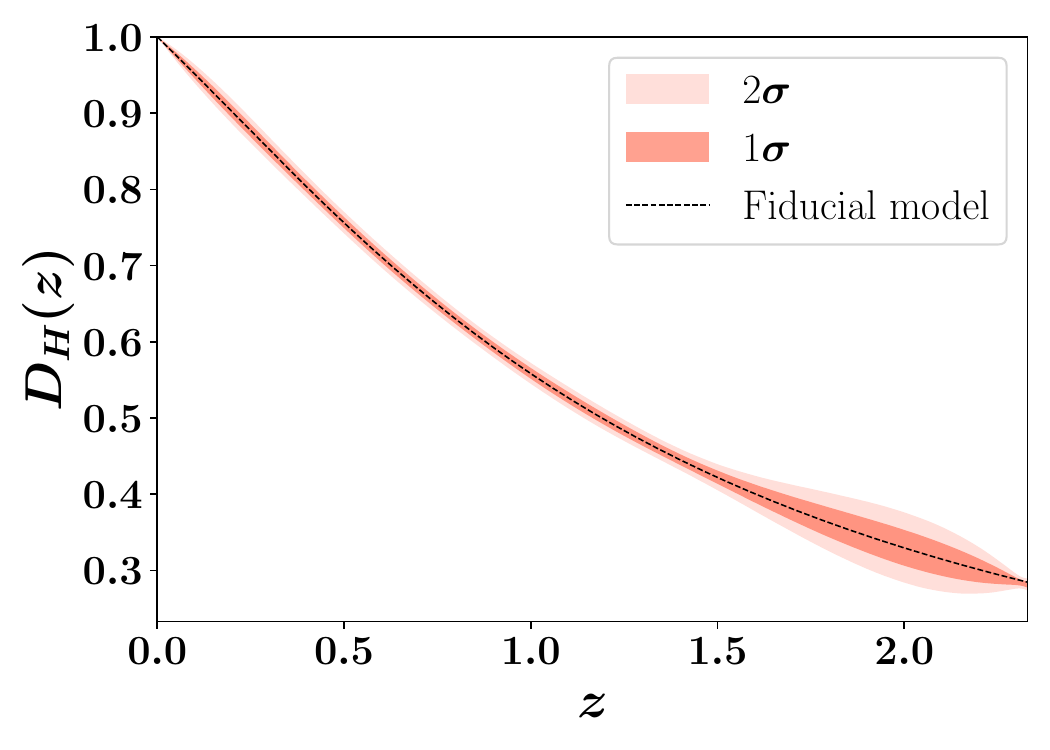}\hfill
\includegraphics[width=0.23\textwidth]{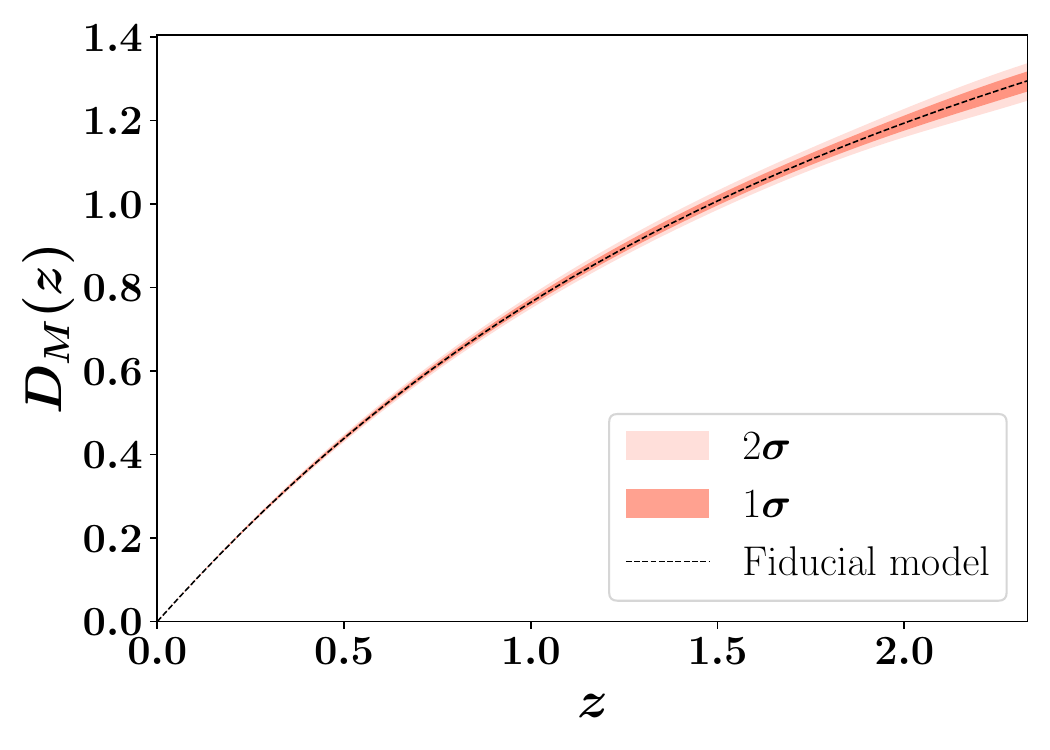}
\includegraphics[width=0.23\textwidth]{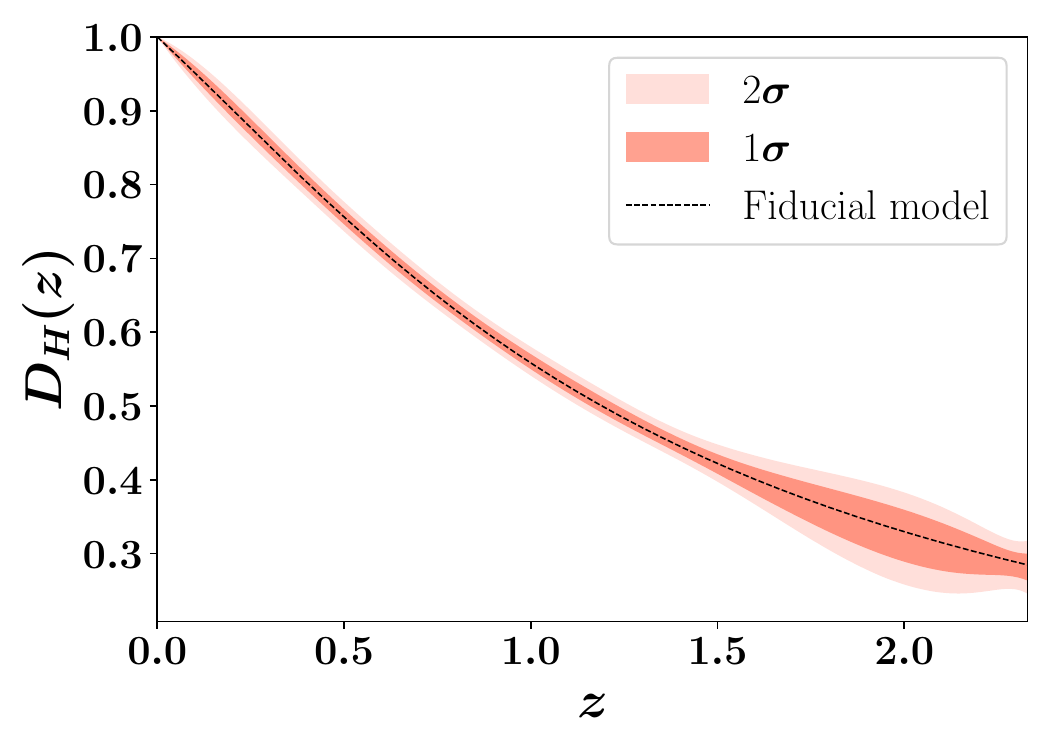}\\
\vspace{0.3cm}
\includegraphics[width=0.23\textwidth]{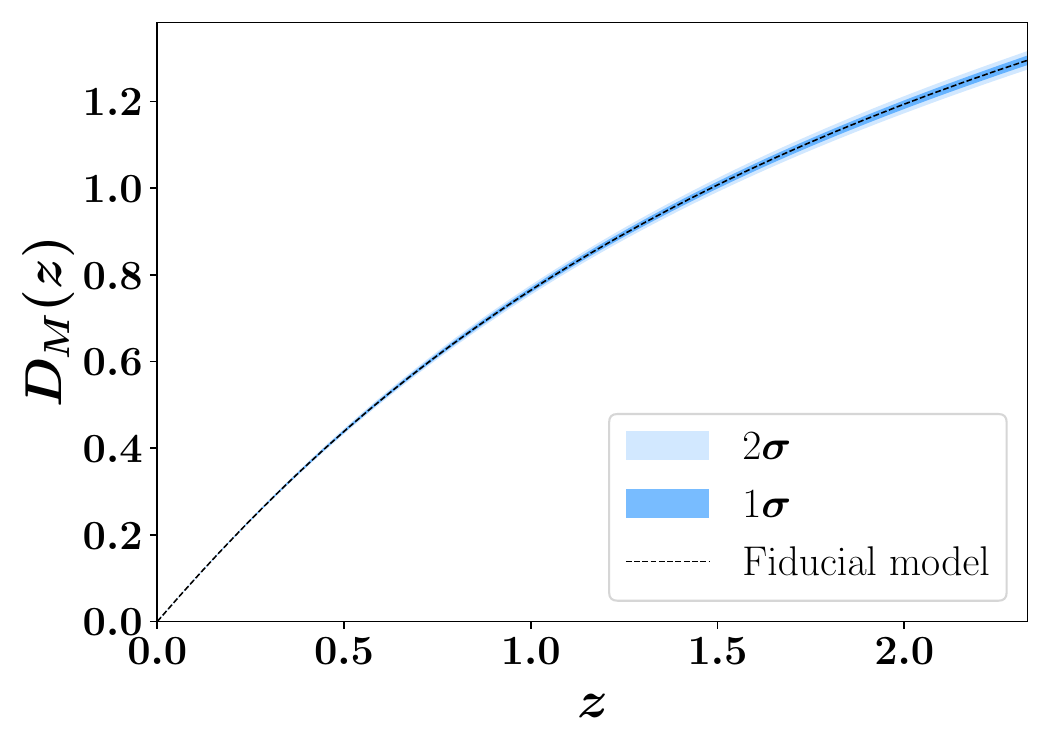}
\includegraphics[width=0.23\textwidth]{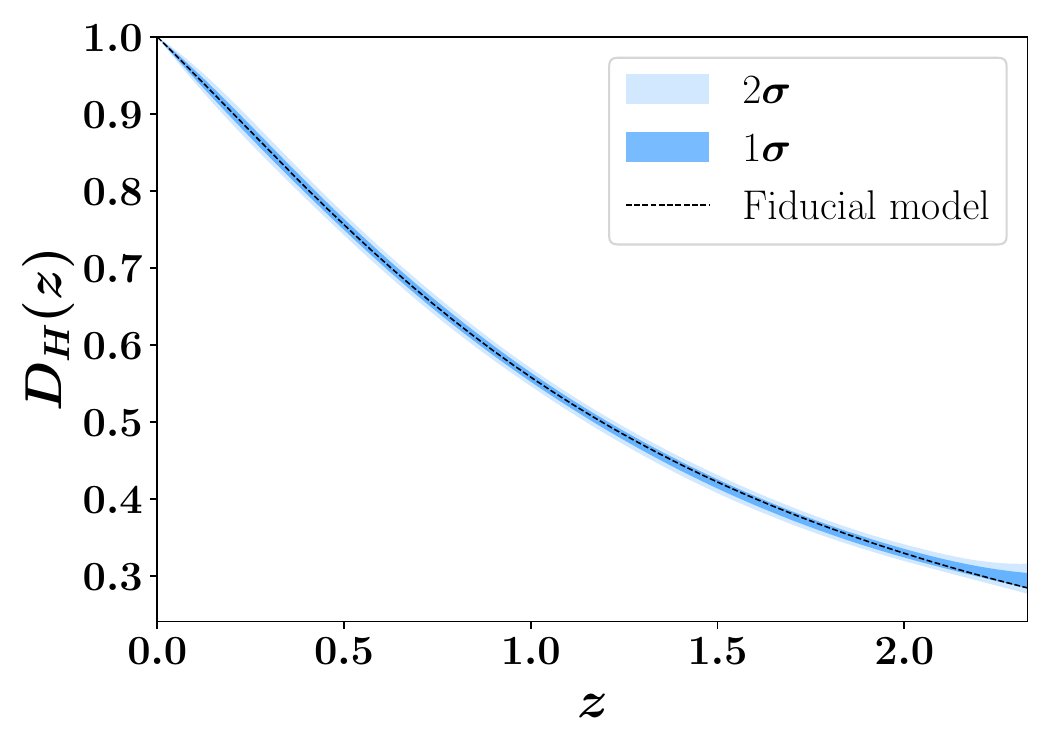}\hfill
\includegraphics[width=0.23\textwidth]{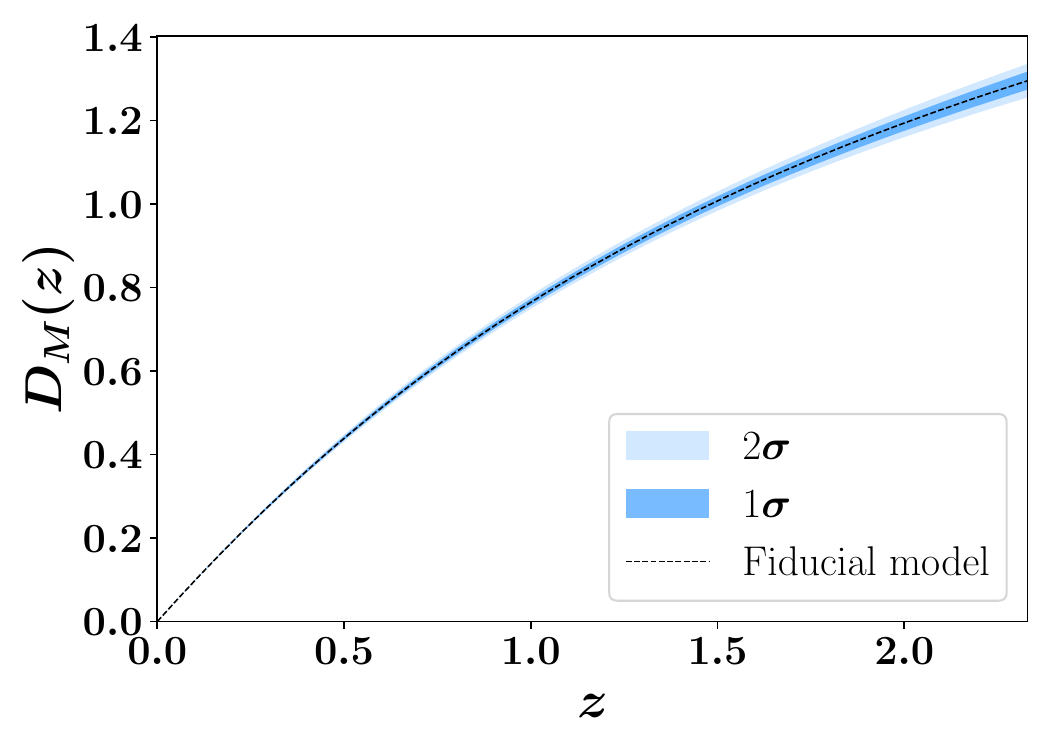}
\includegraphics[width=0.23\textwidth]{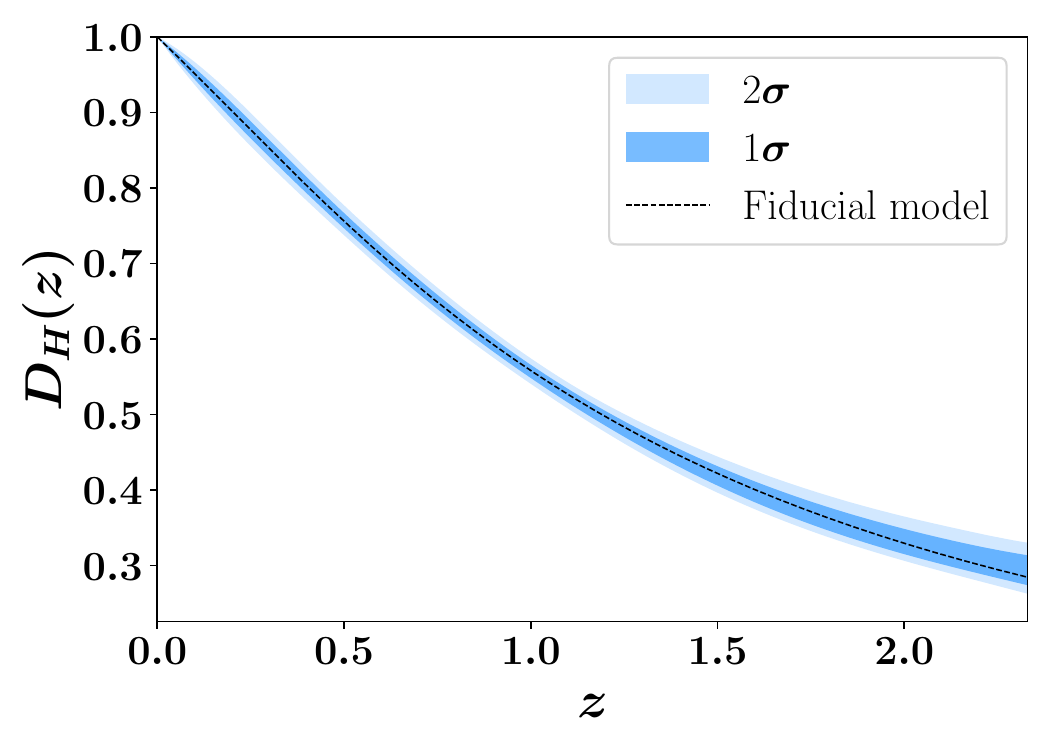}\\
\includegraphics[width=0.45\textwidth]{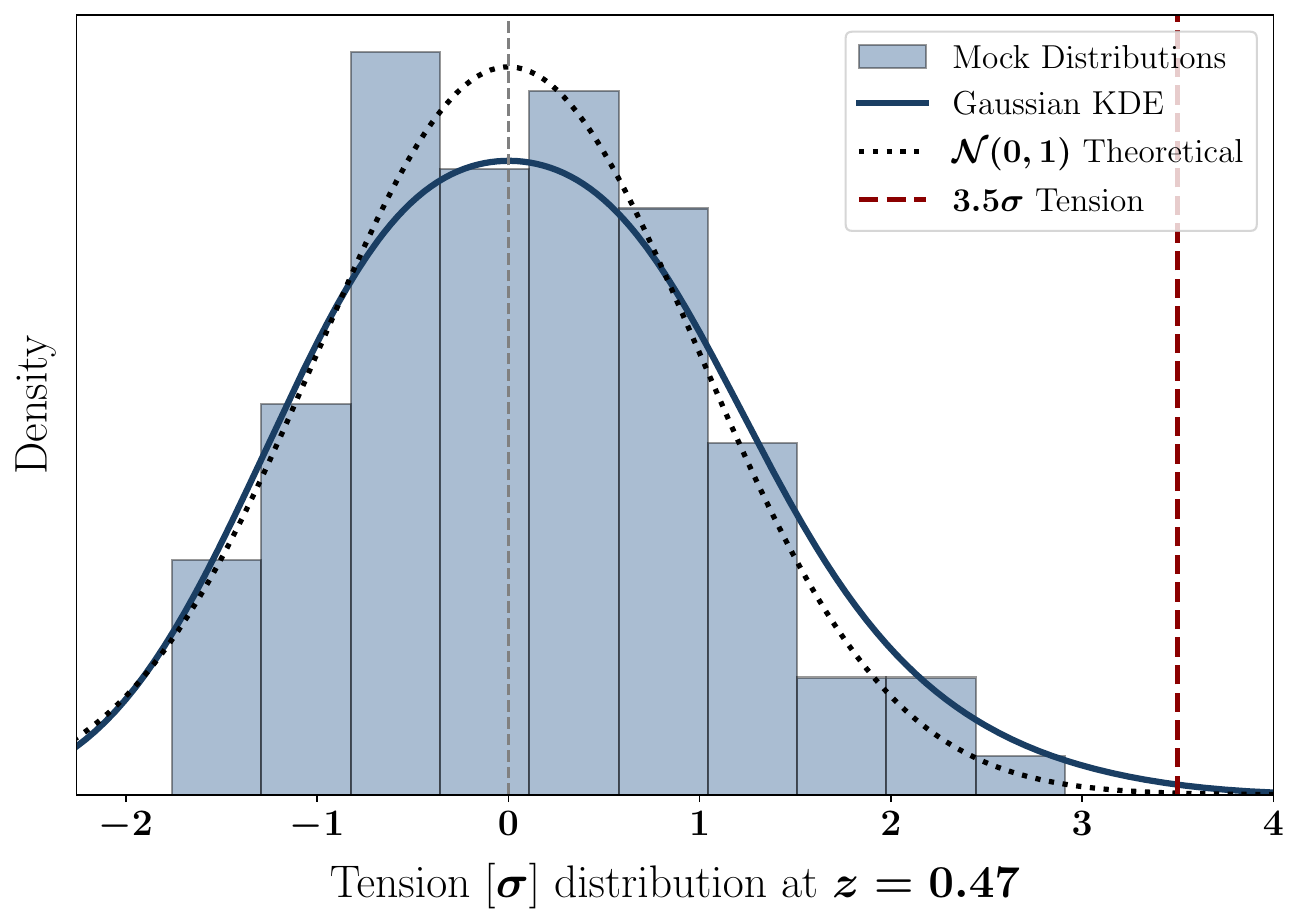}
\caption{
Validation of the reconstruction framework using mock datasets. The upper panels show the RA reconstruction of $D_M(z)$ and $D_H(z)$ for a noiseless mock realization (left) and for the combined posterior samples from 100 noisy mock realizations (right). The middle panels show the corresponding results obtained with REIRA. The lower panel presents the distribution of the reconstructed $H(z)$ tension at $z=0.47$ for the 100 noisy mock realizations relative to the Planck 2018 $\Lambda$CDM prediction.
A kernel density estimate (KDE) of the mock distribution is shown together with the standard normal distribution, $\mathcal{N}(0,1)$ and their agreement confirms the calibration of the statistic, consistent with the validation tests of Table 1. The vertical red dashed line marks the $3.5\sigma$ tension observed in the real data. None of the 100 mock realizations exceeds this value, yielding an empirical $p < 0.01$ independent of any distributional assumption.
}
\label{fig_RA}
\end{figure}

We first consider an idealized noiseless realization, constructed directly from the fiducial model predictions. The resulting RA and REIRA reconstructions are shown in the left-hand panels of Fig.~\ref{fig_RA}, where we compare the recovered dimensionless comoving distance $D_M(z)$ and Hubble distance $D_H(z)$ with the input cosmology. Both reconstruction methods accurately recover the fiducial model within the expected uncertainties, demonstrating that the reconstruction pipeline is unbiased in the absence of statistical noise and validating our numerical implementation.

We then assess the impact of observational uncertainties by generating 100 independent mock realizations of the DESI-DR2 and DES datasets, with random noise drawn from the corresponding covariance matrices. The complete RA and REIRA analyses are repeated for each realization. For every parameter $\theta$, we compute the posterior mean, $\hat{\theta}_i$, and posterior standard deviation, $\sigma_i$, from the $i$-th realization. We then evaluate four complementary diagnostics to quantify the accuracy of the reconstruction, the calibration of the inferred uncertainties, and the statistical significance of the observed anomaly:

\begin{enumerate}

\item {\it Bias test}: We define the normalized bias as
\begin{equation}
B=
\frac{\langle\hat{\theta}\rangle-\theta_{\rm fid}}
{\sigma_{\rm mocks}},
\end{equation}
where $\theta_{\rm fid}$ is the fiducial value, $\langle\hat{\theta}\rangle$ is the average posterior mean over all mock realizations, and $\sigma_{\rm mocks}$ is the scatter of the recovered posterior means. An unbiased reconstruction is expected to satisfy $B\simeq0$.

\item {\it Error calibration test}: We define the error calibration ratio as
\begin{equation}
R=
\frac{\sigma_{\rm mocks}}
{\langle\sigma_{\rm post}\rangle},
\end{equation}
where $\langle\sigma_{\rm post}\rangle$ is the average posterior uncertainty across the mock realizations. A well-calibrated reconstruction is expected to satisfy $R\simeq1$.

\item {\it Pull test}: For each realization, we define the pull statistic as
\begin{equation}
p_i=
\frac{\hat{\theta}_i-\theta_{\rm fid}}
{\sigma_i}.
\end{equation}
For an unbiased reconstruction with correctly estimated uncertainties, the pull distribution is expected to have zero mean and unit variance.

\item {\it Coverage test}: We compute the fraction of mock realizations for which the fiducial value lies within the reconstructed $68\%$ and $95\%$ credible intervals. A well-calibrated reconstruction is expected to recover coverage fractions close to the nominal values of $0.68$ and $0.95$, respectively.

\end{enumerate}

The results are summarized in Table~\ref{tab:mock_validation}. For RA, all reconstructed parameters, with the exception of the final boundary knot at $z=2.33$, successfully pass the bias, pull, uncertainty-calibration, and coverage tests. The normalized biases are consistent with zero, the pull distributions have approximately unit width, and the empirical coverage fractions agree well with the nominal confidence levels. The only significant deviation occurs at the boundary knot, where both the uncertainty-calibration ratio and the pull width increase to approximately $1.87$, while the empirical $68\%$ and $95\%$ coverages decrease to $37\%$ and $72\%$, respectively. This indicates that the uncertainty at the final knot is underestimated by roughly a factor of two. The degradation is expected and arises from the absence of observational constraints between the highest-redshift DES supernova and the final DESI {\bf \texttt{Lya}} measurement at $z=2.33$, making the reconstruction particularly susceptible to boundary effects.

For REIRA, the uncertainty propagation at the final knot improves modestly. The uncertainty-calibration ratio and pull width decrease to approximately $1.66$, while the corresponding $68\%$ and $95\%$ coverages increase to $43\%$ and $74\%$, respectively. Although the performance at $z_5$ deteriorates slightly compared with RA, it remains well within acceptable limits. Aside from these expected boundary effects at the highest-redshift knot, both reconstruction methods are found to be unbiased and to provide reliable uncertainty estimates over the redshift range relevant to this work.

Furthermore, by combining the posterior samples from all 100 noisy mock realizations, we reconstruct the ensemble evolution of $D_M(z)$ and $D_H(z)$ for both the RA and REIRA frameworks. The results, shown in the right-hand panels of Fig.~\ref{fig_RA}, demonstrate that the pipeline consistently recovers the underlying fiducial cosmology within the expected confidence intervals, even in the presence of realistic observational noise.
The lower panel of Fig.~\ref{fig_RA} shows the distribution of the reconstructed $H(z)$ tension at $z=0.47$ for the 100 mock realizations relative to the Planck 2018 $\Lambda$CDM prediction. The distribution is centred around zero, as expected for a fiducial $\Lambda$CDM cosmology, with no realization approaching the $3.5\sigma$ tension observed in the real data. This result indicates that the localized anomaly identified in the DESI-DR2 and DES observations is unlikely to arise from reconstruction bias, statistical fluctuations, or miscalibrated uncertainties under the fiducial $\Lambda$CDM model.

\begin{table*}
\caption{
Validation statistics obtained from 100 independent mock realizations. For each reconstructed parameter, we report the normalized bias,
$(\langle\hat{\theta}\rangle-\theta_{\rm fid})/\sigma_{\rm mocks}$,
the uncertainty-calibration ratio,
$\sigma_{\rm mocks}/\langle\sigma_{\rm post}\rangle$,
the mean and standard deviation of the pull distribution, and the empirical $68\%$ and $95\%$ coverage fractions for both the RA and REIRA reconstructions.
}
\label{tab:mock_validation}

\begin{tabular}{l|c|c|c|c|c|c}
\hline
Parameter &
$(\langle\hat{\theta}\rangle-\theta_{\rm fid})/\sigma_{\rm mocks}$ &
$\sigma_{\rm mocks}/\langle\sigma_{\rm post}\rangle$ &
Pull Mean &
Pull Width &
68\% Coverage &
95\% Coverage \\
\hline
\multicolumn{7}{c}{\textbf{RA}}\\
\hline
$\Delta D_M(z_1)$ & 0.080 & 1.025 & 0.072 & 1.025 & 0.69 & 0.92 \\
$\Delta D_M(z_2)$ & 0.068 & 1.012 & 0.060 & 1.011 & 0.69 & 0.95 \\
$\Delta D_M(z_3)$ & 0.067 & 1.012 & 0.060 & 1.011 & 0.63 & 0.96 \\
$\Delta D_M(z_4)$ & 0.131 & 1.035 & 0.127 & 1.032 & 0.66 & 0.93 \\
$\Delta D_M(z_5)$ & 0.103 & 1.016 & 0.096 & 1.013 & 0.67 & 0.97 \\
$\Delta D_M(z_6)$ & -0.091 & 1.868 & -0.198 & 1.877 & 0.37 & 0.72 \\
$h$ & 0.019 & 1.049 & 0.010 & 1.045 & 0.67 & 0.95 \\
\hline
\multicolumn{7}{c}{\textbf{REIRA}}\\
\hline
$\Delta D_M(z_1)$ & -0.008 & 1.012 & -0.021 & 1.015 & 0.66 & 0.94 \\
$\Delta D_M(z_2)$ & 0.024 & 1.011 & 0.014 & 1.014 & 0.63 & 0.95 \\
$\Delta D_M(z_3)$ & 0.043 & 1.012 & 0.035 & 1.012 & 0.65 & 0.96 \\
$\Delta D_M(z_4)$ & -0.061 & 1.029 & -0.073 & 1.030 & 0.65 & 0.96 \\
$\Delta D_M(z_5)$ & -0.189 & 1.144 & -0.233 & 1.151 & 0.58 & 0.93 \\
$\Delta D_M(z_6)$ & -0.060 & 1.659 & -0.132 & 1.665 & 0.43 & 0.74 \\
$h$ & 0.035 & 1.039 & 0.026 & 1.037 & 0.68 & 0.96 \\
\hline
\end{tabular}
\end{table*}

\end{document}